\documentclass[twocolumn,twocolappendix,times,tighten,trackchanges]{aastex631}

\usepackage{natbib}
\usepackage{xspace}
\usepackage{amsmath}
\usepackage{url}
\usepackage{appendix}

\newcommand{\uat}[2]{\href{http://astrothesaurus.org/uat/#2}{#1 (#2)}}

\newcommand{\lam}{$\lambda$}

\newcommand{\HII}{\mbox{H\,{\sc ii}}}

\newcommand{\SII}{\mbox{S\,{\sc ii}}}

\newcommand{\CII}{\mbox{C\,{\sc ii}}}

\newcommand{\OII}{\mbox{O\,{\sc ii}}}
\newcommand{\OIII}{\mbox{O\,{\sc iii}}}
\newcommand{\NII}{\mbox{N\,{\sc ii}}}

\newcommand{\Ha}{\ensuremath{\mathrm{H\alpha}}}
\newcommand{\Hb}{\ensuremath{\mathrm{H\beta}}}
\newcommand{\Lya}{\ensuremath{\mathrm{Ly\alpha}}}
\newcommand{\CIIfir}{\mbox{[C\,{\sc ii}]\(_{158}\)}}
\newcommand{\OIIIfir}{\mbox{[O\,{\sc iii}]\(_{88}\)}}

\newcommand{\SFR}{\ensuremath{\mathrm{SFR}}}
\newcommand{\SFRUVIR}{\ensuremath{\mathrm{SFR_{UV+IR}}}}

\newcommand{\SFRSED}{\ensuremath{\mathrm{SFR_{SED}}}}
\newcommand{\Zgas}{\ensuremath{Z_\text{gas}}}
\newcommand{\Zsun}{\ensuremath{Z_\text{\(\odot\)}}}

\newcommand{\Mo}{\ensuremath{\mathrm{M_\sun}}}
\newcommand{\Moyr}{\ensuremath{\mathrm{M_\sun\ yr^{-1}}}}

\newcommand{\Lo}{\ensuremath{\mathrm{L_\sun}}}
\newcommand{\LoMoyr}{\ensuremath{\mathrm{L_\sun/M_\sun\ yr^{-1}}}}
\newcommand{\ergs}{\ensuremath{\mathrm{erg\ s^{-1}}}}
\newcommand{\Zo}{\ensuremath{\mathrm{Z_\sun}}}

\newcommand{\um}{$\mathrm{\mu m}$}

\newcommand{\cmmm}{\ensuremath{\mathrm{cm^{-3}}}}
\newcommand{\uG}{\ensuremath{\mathrm{\mu G}}}

\newcommand{\NO}{\ensuremath{\text{N}/\text{O}}}

\newcommand{\OH}{\ensuremath{\text{O}/\text{H}}}

\newcommand{\CO}{\ensuremath{\text{C}/\text{O}}}

\newcommand{\LOiii}{\ensuremath{L(\OIIIfir)}}
\newcommand{\LCii}{\ensuremath{L(\CIIfir)}}

\newcommand{\CPDR}{\ensuremath{C_\text{PDR}}}

\newcommand{\nn}{\mbox{--}}

\shorttitle{Bridging Optical and FIR Emission-Line Diagrams from \(z \sim 0\) to \(> 6\)}
\shortauthors{Sugahara et al.}

\begin{document}

\title{\Large Bridging Optical and Far-Infrared Emission-Line Diagrams of Galaxies from Local to the Epoch of Reionization: Characteristic High \([\OIII] 88 \text{\um} /\SFR\) at \( z > 6\)}

\author[0000-0001-6958-7856]{Yuma Sugahara}
\email{sugayu@aoni.waseda.jp}
\affil{National Astronomical Observatory of Japan, 2-21-1 Osawa, Mitaka, Tokyo 181-8588, Japan}
\affil{Waseda Research Institute for Science and Engineering, Faculty of Science and Engineering, Waseda University, 3-4-1, Okubo, Shinjuku, Tokyo 169-8555, Japan}
\author[0000-0002-7779-8677]{Akio K. Inoue}
\affil{Waseda Research Institute for Science and Engineering, Faculty of Science and Engineering, Waseda University, 3-4-1, Okubo, Shinjuku, Tokyo 169-8555, Japan}
\affil{Department of Physics, School of Advanced Science and Engineering, Faculty of Science and Engineering, Waseda University, 3-4-1, Okubo, Shinjuku, Tokyo 169-8555, Japan}
\author[0000-0001-7440-8832]{Yoshinobu Fudamoto}
\affil{National Astronomical Observatory of Japan, 2-21-1 Osawa, Mitaka, Tokyo 181-8588, Japan}
\affil{Waseda Research Institute for Science and Engineering, Faculty of Science and Engineering, Waseda University, 3-4-1, Okubo, Shinjuku, Tokyo 169-8555, Japan}
\author[0000-0002-0898-4038]{Takuya Hashimoto}
\affil{Tomonaga Center for the History of the Universe (TCHoU), Faculty of Pure and Applied Sciences, University of Tsukuba, Tsukuba, Ibaraki 305-8571, Japan}
\author[0000-0002-6047-430X]{Yuichi Harikane}
\affil{Institute for Cosmic Ray Research, The University of Tokyo, 5-1-5 Kashiwanoha, Kashiwa, Chiba 277-8582, Japan}
\affil{Department of Physics and Astronomy, University College London, Gower Street, London WC1E 6BT, UK}
\author[0000-0002-7738-5290]{Satoshi Yamanaka}
\affil{General Education Department, National Institute of Technology, Toba College, 1-1, Ikegami-cho, Toba, Mie 517-8501, Japan}

\begin{abstract}
We present photoionization modeling of galaxy populations at \(z\sim0\), \(2\), and \( > 6\) to bridge optical and far-infrared (FIR) emission-line diagrams.
We collect galaxies with measurements of optical and/or FIR ([\OIII] 88 \um\ and [\CII] 158 \um) emission line fluxes and plot them on the \([\OIII]\lambda5007/\Hb\)--\([\NII]\lambda6585/\Ha\) (BPT) and \(\LOiii/\SFR\)--\(\LCii/\SFR\) diagrams, where \SFR\ is the star-formation rate and \LOiii\ and \LCii\ are the FIR line luminosities.
We aim to explain the galaxy distributions on the two diagrams with photoionization models that employ three nebular parameters: the ionization parameter \(U\), hydrogen density \(n_\text{H}\), and gaseous metallicity \Zgas.
Our models successfully reproduce the nebular parameters of local galaxies, and then predict the distributions of the \(z\sim0\), \(2\), and \( > 6\) galaxies on the diagrams.
The predicted distributions illustrate the redshift evolution on all the diagrams; e.g., \([\OIII]/\Hb\) and \(\OIIIfir/\CIIfir\) ratios continuously decrease from \(z > 6\) to \(0\).
Specifically, the \(z > 6\) galaxies exhibit \(\sim\!0.5\) dex higher \(U\) than low-redshift galaxies at a given \Zgas\ and show predicted flat distributions on the BPT diagram at \(\log{[\OIII]/\Hb} = 0.5\nn0.8\).
We find that some of the \(z > 6\) galaxies exhibit high \(\LOiii/\SFR\) ratios.
To explain these high ratios, our photoionization models require a low stellar-to-gaseous metallicity ratio or bursty/increasing star-formation history at \(z > 6\).
The \textit{James Webb Space Telescope} will test the predictions and scenarios for the \(z > 6\) galaxies proposed by our photoionization modeling.
\end{abstract}


\section{Introduction}
\label{sec:introduction}

Emission-line ratios reflect physical properties of the interstellar medium (ISM) of galaxies.
In the local universe, optical to far-infrared (FIR) spectroscopy has provided various emission-line measurements to probe essential ISM properties, including the electron temperature, electron density, ionization state, and elemental abundances.
However, direct comparisons of the ISM properties at different redshifts are challenging due to limited sensitivity and wavelength coverage of observational instruments.

Recent near-infrared (NIR) observations have significantly deepened our understanding of the stellar and ISM properties of galaxies at redshift up to \(z\sim2\).
Compared with local galaxies, \(z\sim2\) galaxies have lower metallicities \citep[e.g.,][]{Erb:2006a, Sanders.R:2020a}, higher electron densities \citep[e.g.,][]{Masters.D:2014a, Steidel:2014, Shimakawa.R:2015b, Sanders.R:2016a}, higher ionization parameters \citep[e.g.,][]{Nakajima:2014, Steidel:2016, Trainor.R:2016a, Kashino.D:2017a, Kojima.T:2017a}, harder ionizing spectra \citep[e.g.,][]{Steidel:2014, Steidel:2016, Trainor.R:2016a, Shapley.A:2019a}, and possibly higher nitrogen-to-oxygen (\NO) abundance ratios \citep[e.g.,][]{Masters.D:2014a, Shapley.A:2015a, Kojima.T:2017a}.
Combinations of the redshift evolution would offset galaxy distributions on the Baldwin--Phillips--Terlevich \citep[BPT;][]{Baldwin.J:1981a,Veilleux.S:1987a} diagram \citep[e.g.,][]{Kewley.L:2013b, Bian.F:2020a}, even though the origin of the offset is under debate.
At higher redshift of \(z > 4\), observations of the rest-frame optical emission lines become more difficult because they fall into mid-infrared (MIR) wavelengths.

The Atacama Large Millimeter/submillimeter Array (ALMA) enabled to investigate the ISM properties of high-redshift galaxies, even at \( z > 6\), by observing atomic/ionic fine-structure emission lines in FIR wavelengths.
Popular lines among high-redshift ALMA observations are [\CII] 158 \um\ \citep[hereafter \CIIfir;][]{Capak:2015, Maiolino.R:2015a} and [\OIII] 88 \um\ \citep[\OIIIfir;][]{Inoue.A:2014a, Inoue.A:2016a}, which are the strongest FIR emission lines, and thus important coolants of the ISM \citep{Tielens.A:1985a, Cormier.D:2015a}.
High-redshift galaxies observed with ALMA can be directly compared with local galaxies observed with FIR telescopes like the \textit{Infrared Space Observatory} \citep[\textit{ISO};][]{Kessler.M:1996a} and the \textit{Herschel Space Observatory} \citep{Pilbratt.G:2010a}.
After \citet{Inoue.A:2016a} reported higher \(\OIIIfir/\CIIfir\) line ratios of SXDF-NB1006-2 at \(z = 7.21\) than those of local galaxies, following ALMA observations supported high \(\OIIIfir/\CIIfir\) ratios among \( z > 6\) galaxies \citep[e.g.,][but see \citealp{Carniani.S:2020a}]{Hashimoto.T:2019a, Harikane.Y:2020b, Bakx.T:2020a}.
These high \(\OIIIfir/\CIIfir\) ratios invoke the ISM and star-formation properties characteristic of high-redshift galaxies; for example, high ionization parameters \citep[][]{Harikane.Y:2020b}, low covering fractions of a photo-dissociation region (PDR) surrounding \HII\ regions \citep[][]{Harikane.Y:2020b}, bursty star-formation history \citep{Arata.S:2020a, Vallini.L:2021a}, low carbon-to-oxygen (\CO) abundance ratios \citep{Arata.S:2020a, Katz.H:2022a}, and top-heavy initial mass functions \citep[IMF;][]{Katz.H:2022a}.

Comparisons with local galaxies have unveiled the stellar and ISM properties of galaxies at \(z\sim2\) and \(z > 6\); however, galaxies in the two high-redshift ranges cannot be directly compared with each other because no emission lines are commonly observed for the two populations.
One of the key tools to overcome this difficulty is photoionization models.
Photoionization models simplify ISM structures of galaxies under some assumptions and predict emission-line intensity ratios from an input ionizing spectrum and nebular physical parameters.
They are widely adopted to galaxy population at each redshift, including local galaxies like dwarf galaxies \citep{Cormier.D:2019a} and (ultra-)luminous infrared galaxies \citep[U/LIRGs; e.g.,][]{Nagao.T:2011a, Inami.H:2013a, Pereira-Santaella.M:2017a} and \(z\sim2\) galaxies \citep[e.g.,][]{Steidel:2014, Sanders.R:2016a, Trainor.R:2016a, Strom.A:2018a}.
For \(z > 6\) galaxies, \citet{Harikane.Y:2020b} used photoionization models to propose a \(L(\OIIIfir)/\SFR\)--\(L(\CIIfir)/\SFR\) diagram, where \LOiii\ and \LCii\ are the \OIIIfir\ and \CIIfir\ line luminosities, respectively, and \SFR\ is the star-formation rate.
They concluded that high-redshift galaxies have high ionization parameter, low PDR covering fraction, or both.
By applying a similar model to Harikane et al., \citet{Sugahara.Y:2021a} constrained the \NO\ ratio of B14-65666 at \(z = 7.15\) as a function of the metallicity.

In this paper we attempt to bridge local to \(z > 6\) galaxy populations by modeling their distributions on diagrams of various emission-line ratios.
Our attempt derives average ISM properties of these galaxy populations and infers any evolutionary trends as a function of redshift.
This paper is organized as follows.
Section \ref{sec:observational-data} describes samples of galaxy populations at \(z\sim0\), \(2\), and \( > 6\) that have optical and/or FIR emission-line measurements.
Section \ref{sec:photo-ioniz-model} explains setups of photoionization models.
Section \ref{sec:results} shows results of photoionization modeling of galaxy populations in emission-line diagrams.
Section \ref{sec:discussion} discusses implications of line ratios, parameter dependence of photoionization models, and caveats.
Finally Section \ref{sec:summary} summarizes our findings.
Throughout this paper, we use the \citet{Chabrier:2003} initial mass function (IMF) in the mass range of \(0.1\)--\(100\) \Mo\ and convert \SFR\ from \citet{Salpeter:1955} and \citet{Kroupa:2001} IMFs by multipying factors of \(0.63\) and \(0.94 ( = 0.63/0.67)\), respectively \citep{Madau:2014}.

\section{Observational Data}
\label{sec:observational-data}

We aim to model distributions of galaxy populations on two emission-line diagrams: \([\OIII]/\Hb\nn[\NII]/\Ha\) and \(\LOiii/\SFR \nn \LCii/\SFR\) diagrams.
In this paper the two diagrams are referred to as the BPT and FIR diagrams, respectively.
We collected galaxies with emission-line measurements that can be plotted on the BPT and FIR diagrams.
Section \ref{sec:optical-fir-samples} constructs a sample of local galaxies with both optical and FIR line measurements.
Section \ref{sec:optical-samples} constructs a statistical sample of galaxies at \(z \sim 0\) and \(2\) with only optical line measurements.
Section \ref{sec:fir-samples} constructs a sample of \( z > 6 \) galaxies observed with ALMA.

\begin{figure*}[t]
    \epsscale{1.1}
    \plottwo{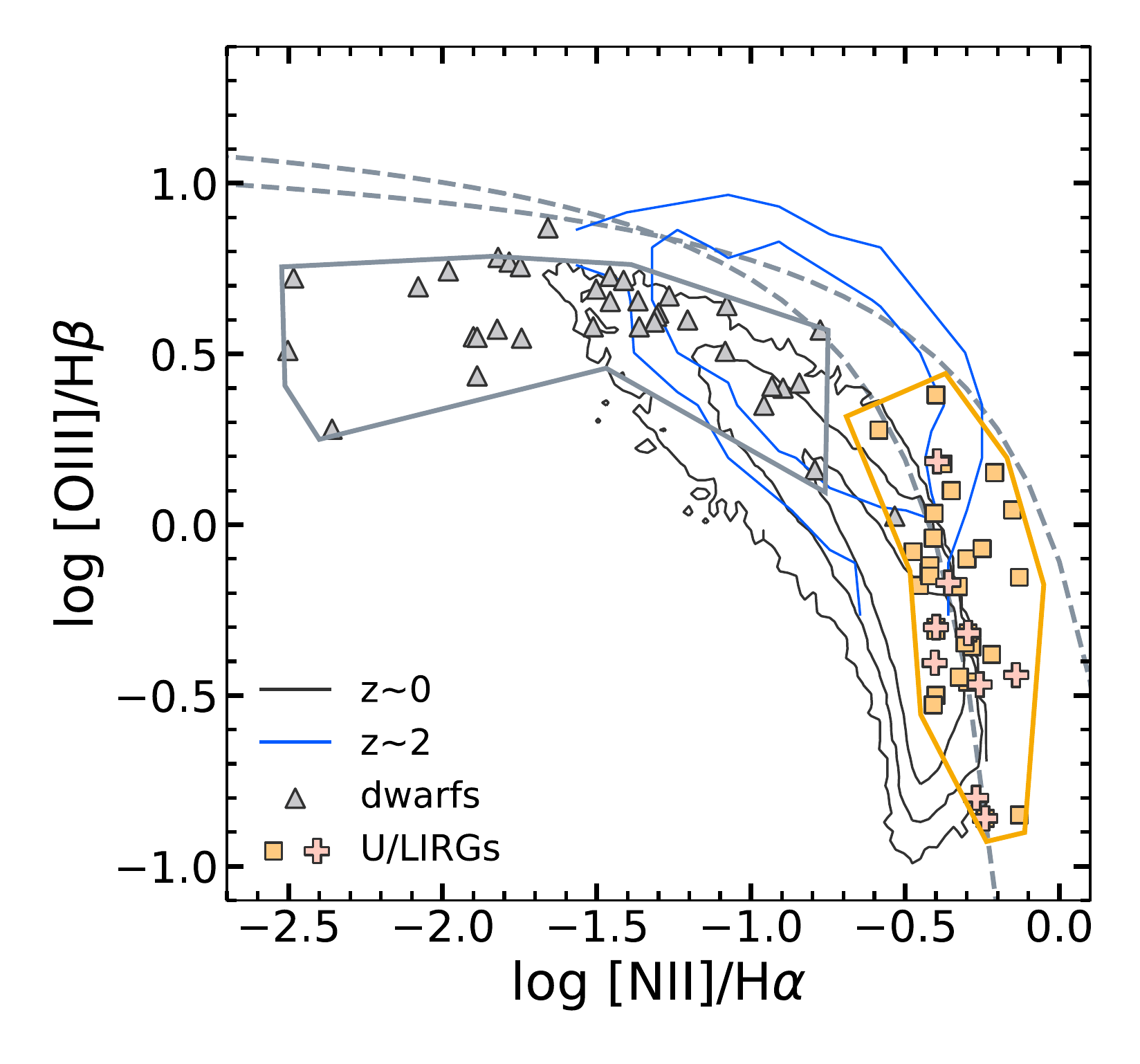}{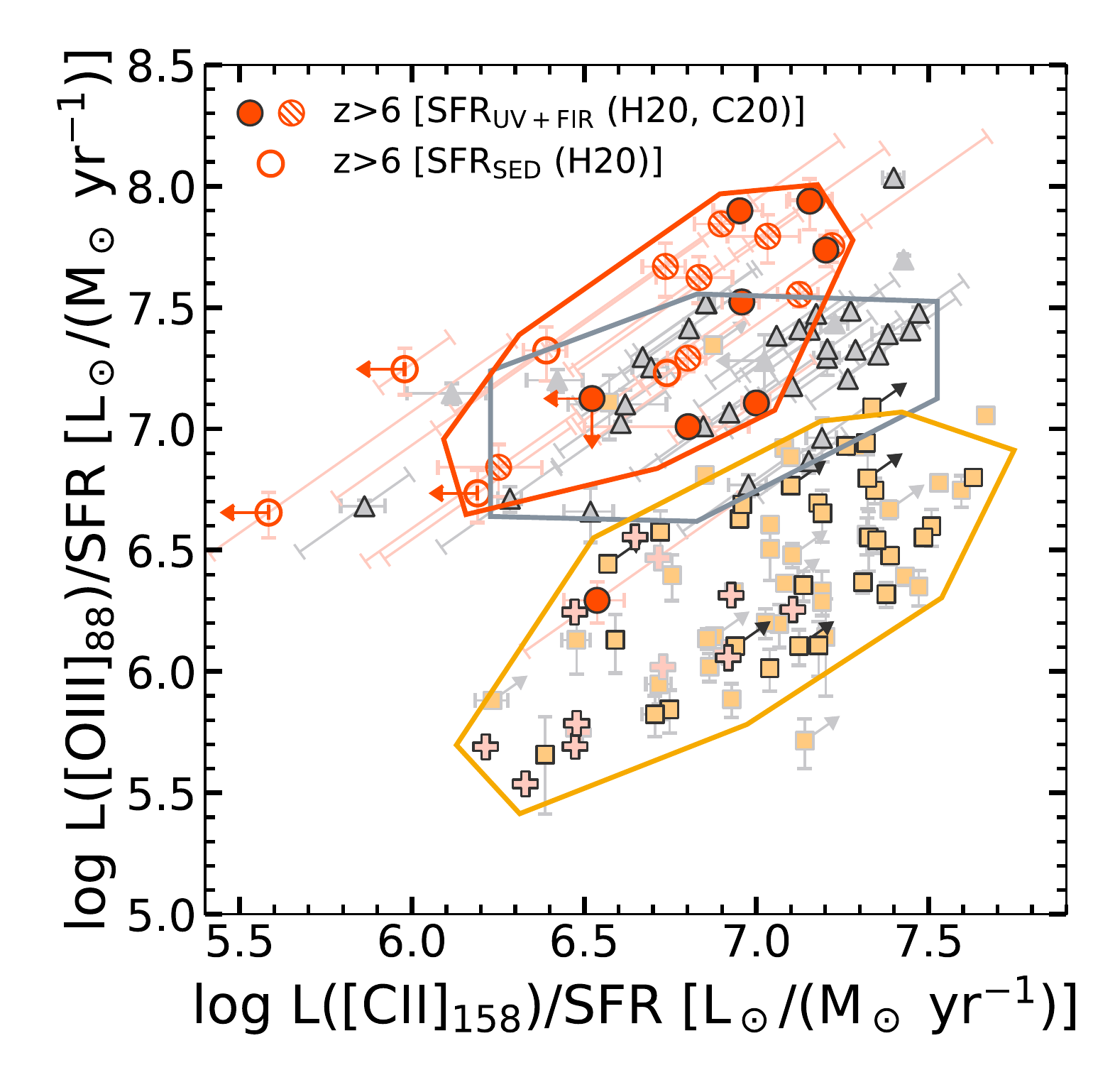}
    \caption{BPT diagram (Left) and FIR diagram (Right).
      The red circles show the \(z > 6\) galaxies; the filled circles show the measurements mainly taken from \citet{Harikane.Y:2020b}, while the hashed circles from \citet{Carniani.S:2020a}.
      The open red circles are taken from \citet{Harikane.Y:2020b}, but their \SFR\ is measured with the SED fitting.
      The gray triangles show the local dwarfs \citep{Cormier.D:2019a} and the orange symbols show the local U/LIRGs including the GOALS galaxies \citep[square,][]{Diaz-Santos.T:2017a} and the SHINING galaxies \citep[cross,][]{Herrera-Camus.R:2018a}.
      The solid polygons depict the regions representing the distributions of the galaxy populations, which are used in our photoionization modeling.
      \textit{Left}: The black and blue contours illustrate the distributions of the \(z\sim0\) (SDSS) and \(z \sim 2\) \citep{Strom.A:2017b, Shivaei.I:2018a} galaxies, respectively, which include 68, 95, and 99.5\% of the \(z\sim0\) galaxies and 68 and 90\% of the \(z \sim 2\) galaxies.
      The dashed gray lines depict the criteria between star-forming galaxies and AGNs \citep{Kauffmann.G:2003a, Kewley.L:2001b}.
      \textit{Right}: The horizontal and vertical error bars indicate the \CIIfir\ and \OIIIfir\ measurement errors, respectively, and the diagonal error bars does the \SFR\ errors.
      For local galaxies, the black edges mean the data points with optical measurements, while the gray edges mean ones without them.
    }
  \label{fig:bptfir}
\end{figure*}

\subsection{OptFIR sample: local dwarfs and U/LIRGs}
\label{sec:optical-fir-samples}
The first sample consists of local galaxies with both optical (\Hb\ \lam4861, [\OIII] \lam5007, \Ha\ \lam6563, and [\NII] \lam6585) and FIR (\OIIIfir\ and \CIIfir) line measurements, being referred to as the optFIR sample.
We firstly describe FIR observations, with the PACS instrument \citep{Poglitsch.A:2010a} on board \textit{Herschel}, and then explain optical line measurements.
This sample is mainly divided into two galaxy populations: local dwarfs and local U/LIRGs.

The local dwarfs were taken from the Dwarf Galaxy Survey \citep[DGS, PI: Madden;][]{Madden.S:2013a, Cormier.D:2015a}, which is targeting 50 dwarf galaxies.
We used 36 galaxies for which \citet{Cormier.D:2015a} listed both \CIIfir\ and \OIIIfir\ line fluxes.
\SFR\ was taken from \citet{De-Looze.I:2014a} and \citet{Madden.S:2013a}; De Looze et al.\! derived \SFR\ from \textit{GALEX} far-ultraviolet (FUV) and \textit{Spitzer}/MIPS 24 \um\ fluxes using formulae in \citet{Hao.C:2011a} and \citet{Murphy.E:2011a}, whereas Madden et al.\! derived it from total infrared luminosity (\(L_\text{TIR}\)) or \Ha\ (or \Hb) luminosity.
Here, values from De Looze et al.\! have priority if galaxies are listed in both references.
The stellar mass spans from \(\log{M_{*}/\Mo} = 6.5\) to \(10.5\) with a mean of \(8.6\) and a standard deviation of \(1.0\) dex \citep{Madden.S:2013a}.

The local U/LIRGs were composed of 61 galaxies taken from the Great Observatories All-Sky LIRG Survey \citep[GOALS;][]{Armus.L:2009a, Diaz-Santos.T:2013a} and 14 galaxies taken from the Survey of Far-infrared Lines in Nearby Galaxies \citep[SHINING, PI: Strum;][]{Herrera-Camus.R:2018a, Herrera-Camus.R:2018b}.
The selected galaxies have both \CIIfir\ and \OIIIfir\ line flux measurements.
We excluded from the sample active galactic nuclei (AGNs) classified in the literature \citep{Rich.J:2015a, Herrera-Camus.R:2018a}, from the bolometric AGN fractional contribution of \( \langle \alpha_\text{AGN}^\text{bol} \rangle \geq 0.5  \) \citep{Diaz-Santos.T:2017a}, and on the BPT diagram \citep{Kewley.L:2001b}.
For the GOALS galaxies, the line luminosities were converted from ``Best line flux'' given by \citet{Diaz-Santos.T:2017a} using the luminosity distance in \citet{Armus.L:2009a}.
\SFR\ was taken from \citet{Howell.J:2010a}, who computed it from \textit{GALEX} FUV and \textit{IRAS} \(L_\text{TIR}\) \citep{Kennicutt.R:1998a}.
For the SHINING galaxies, the line luminosities were converted from ``integrated line fluxes'' given by \citet{Herrera-Camus.R:2018a} using given redshifts.
\SFR\ was computed from \(L_\text{TIR}\) \citep{Murphy.E:2011a}, where \(L_\text{TIR} = 1.75 L_\text{FIR}\) \citep{Herrera-Camus.R:2018a} and \(L_\text{FIR}\) is the FIR luminosity at \(42.5\)--\(122.5\) \um\ \citep{Helou.G:1988a}.
The stellar mass is \(\log{M_{*}/\Mo} \simeq 10.6\nn11.6\) with a mean of \(11.1\) for the GOALS galaxies \citep{Howell.J:2010a}, which is similar to the mass of the SHINING galaxies \citep{Herrera-Camus.R:2018b}.
We note that the U/LIRGs defined here include galaxies that do not satisfy the definition of LIRGs, \(L_\text{IR} > 10^{11}\) \Lo, but we categorized them as U/LIRGs in this work for simplicity.

Optical emission-line ratios, \([\OIII]/\Hb\) and \([\NII]/\Ha\), are taken from the literature \citep{Veilleux.S:1999a, Kewley.L:2001a, Moustakas.J:2006a, Moustakas.J:2010a, Rich.J:2015a, De-Vis.P:2017a, Perna.M:2021a}\footnote{As there is no table listing line ratios in \citet{Rich.J:2015a}, we took typical values from distributions of spaxels on the BPT diagram. The uncertainties caused by this method do not affect our results.}.
When a galaxy is listed in multiple studies, the measurements from the integrated spectroscopic observations \citep{Moustakas.J:2006a, Moustakas.J:2010a, Rich.J:2015a, Perna.M:2021a} have priority, considering a consistency with a large field of view of FIR observations.
In this way, optical measurements were taken for 32 out of the DGS galaxies, 29 out of the GOALS galaxies, and 9 out of the SHINING galaxies.

Figure \ref{fig:bptfir} shows the distributions of the optFIR sample on the BPT and FIR diagrams.
The dwarfs exhibit higher \([\OIII]/\Hb\) and \(\LOiii/\SFR\) values than the U/LIRGs, suggesting higher ionization states and lower metallicities of the dwarfs.
The low \(\LCii/\SFR\) and \(\LOiii/\SFR\) ratios in the U/LIRGs are known as ``line deficit'' \citep{Diaz-Santos.T:2017a, Herrera-Camus.R:2018a, Herrera-Camus.R:2018b}, lower line-to-\(L_\text{FIR}\) ratios at higher \(L_\text{FIR}\) or higher \SFR.

\subsection{Optical sample: \( z \sim 0\) and \(2\) galaxies}
\label{sec:optical-samples}

The second sample consists of galaxies at \(z\sim0\) and \(2\) that have measurements of the optical \([\OIII]/\Hb\) and \([\NII]/\Ha\) ratios but not the FIR observations.
We refer to this sample as the optical sample.

The \(z\sim0\) galaxies were drawn from the Sloan Digital Sky Survey \citep[SDSS;][]{York:2000} Data Release 7 \citep{Abazajian:2009} main galaxy sample \citep{Strauss:2002}.
The emission-line fluxes are given in the MPA-JHU catalog\footnote{https://wwwmpa.mpa-garching.mpg.de/SDSS/DR7}.
We selected \(\sim\)20,000 star-forming or starburst galaxies detected in \Hb, [\OIII], [\NII], and \Ha\ lines with \(\mathrm{S/N} > 5\).
Composites of AGNs and starbursts were removed based on the criterion of \citet{Kauffmann.G:2003a} on the BPT diagram.
The mean and standard deviation of the stellar mass taken from the same catalog are \(\log{M_{*}/\Mo} = 10.0\) and \(0.7\) dex, respectively \citep{Kauffmann.G:2003a}.

The \(z\sim2\) galaxies consisted of Keck Baryon Structure Survey observed with MOSFIRE \citep[KBSS-MOSFIRE;][]{Steidel:2014} and the MOSFIRE Deep Evolution Field survey \citep[MOSDEF;][]{Kriek:2015}.
We took the emission-line ratios of 226 galaxies from \citet{Strom.A:2017b} for the KBSS-MOSFIRE survey and those of 223 galaxies from \citet{Shivaei.I:2018a} for the MOSDEF survey.
Upper limits of the line ratios were not included in the sample.
We also did not include objects as AGN candidates whose \([\OIII]/\Hb\) values are \(> 0.2\) dex higher than the criterion of \citet{Kewley.L:2001b}.
The stellar mass for the KBSS-MOSFIRE galaxies spans from \(\log{M_{*}/\Mo} \simeq 8.6\) to \(11.4\) with a mean of \(10.0\) \citep{Steidel:2014} and the stellar mass for the MOSDEF galaxies similarly spans from \(9.0\) to \(11.5\) \citep[MOSDEF;][]{Kriek:2015}.

The contours in the left panel of Figure \ref{fig:bptfir} illustrate the distributions of the \(z\sim0\) and \(2\) galaxies on the BPT diagram.
The black contours include 68, 95, and 99.5\% of the \(z\sim0\) galaxies, while the blue contours include 68 and 90\% of the \(z\sim2\) galaxies.
The figure shows a clear offset of the distributions between the \(z\sim0\) and \(2\) galaxies, indicating the evolution of ISM properties as described in Section \ref{sec:introduction}.

\subsection{FIR sample: \( z > 6 \) galaxies}
\label{sec:fir-samples}

The third sample, referred to as the FIR sample, contains galaxies at \(z > 6\) with both ALMA \OIIIfir\ and \CIIfir\ line measurements, including non-detections.
In contrast to abundant \CIIfir\ observations, limited \OIIIfir\ observations determine the sample size of \(\sim\)10 galaxies \citep{Inoue.A:2016a, Carniani.S:2017b, Laporte.N:2017a, Marrone.D:2018a, Hashimoto.T:2018a, Walter.F:2018a, Hashimoto.T:2019a, Tamura.Y:2019a, Harikane.Y:2020b}.

The \(z > 6\) galaxies are provided by two compilations, \citet{Harikane.Y:2020b} and \citet{Carniani.S:2020a}.
\citet{Harikane.Y:2020b} summarized \CIIfir\ and \OIIIfir\ line luminosities and \SFR\ of nine UV-selected galaxies (Lyman-break galaxies and Lyman-alpha emitters) and three submillimeter-galaxies (SMGs) from the literature.
\citet{Carniani.S:2020a} reanalyzed ALMA data of the nine UV-selected galaxies with \textit{uv}-tapering to report \CIIfir\ detections for the galaxies in which previous studies did not detect.
Carniani et al.\! also presented \(\SFRUVIR\), \SFR\ computed from the sum of the UV and IR luminosity, \(L_\text{UV+IR}\) \citep{Kennicutt.R:2012a}.
In this work, we used the \CIIfir\ and \OIIIfir\ line luminosities and \SFR\ from the both compilations.
For J1211-0118, J0235-0532, and J0217-0208 \citep{Harikane.Y:2020b}, we used the \SFR\ values updated by Ono et al.\! in prep.
Moreover, we added z7\_GSD\_3811 \citep{Binggeli.C:2021a} to our sample.
Although deriving the stellar mass is difficult for the \(z > 6\) galaxies, it was inferred from the spectral-energy-distribution (SED) fitting and empirical relations of UV photometry in the literature.
The inferred stellar mass spans from \(\log{M_{*}/\Mo} \sim 9\) to \(10.5\) \citep{Inoue.A:2016a, Laporte.N:2017a, Marrone.D:2018a, Hashimoto.T:2018a, Hashimoto.T:2019a, Tamura.Y:2019a, Harikane.Y:2020b, Binggeli.C:2021a}.

The sample finally consists of 13 galaxies with 22 data points.
In the right panel of Figure \ref{fig:bptfir}, the red circles show the \(z > 6\) galaxies on the FIR diagram.
Regarding the galaxies that have different measurements from the two compilations, the filled red circles show measurements for \citet{Harikane.Y:2020b} and the hashed red circles for \citet{Carniani.S:2020a}.
The open red circles depict the galaxies whose \SFR\ is computed from SED fitting, \SFRSED.
We mainly analyze galaxies with \SFRUVIR\ in this work.
For convenience, Figure \ref{fig:fir-name} corresponds the names of the \(z > 6\) galaxies to the data points on the FIR diagrams.

The \(\LOiii/\SFR\) ratios of the \(z > 6\) galaxies are comparable to or higher than the local dwarfs, despite their high \SFR.
These high \(\LOiii/\SFR\) ratios at a given \SFR\ are consistent with previous results of higher line-to-\(L_\text{FIR}\) ratios at \(z > 1\) than local line deficit at a given \(L_\text{FIR}\) \citep[e.g.,][]{Herrera-Camus.R:2018a, Sugahara.Y:2021a}.
The three SMGs exhibit lower \(\LOiii/\SFR\) than the other \(z > 6\) galaxies at a given \(\LCii/\SFR\), implying that SMGs would be analogues of local U/LIRGs, which exhibit lower \(\LOiii/\SFR\) than local dwarfs.

\section{Fiducial Photoionization Models}
\label{sec:photo-ioniz-model}

\begin{figure*}[t]
    \epsscale{1.1}
    \plottwo{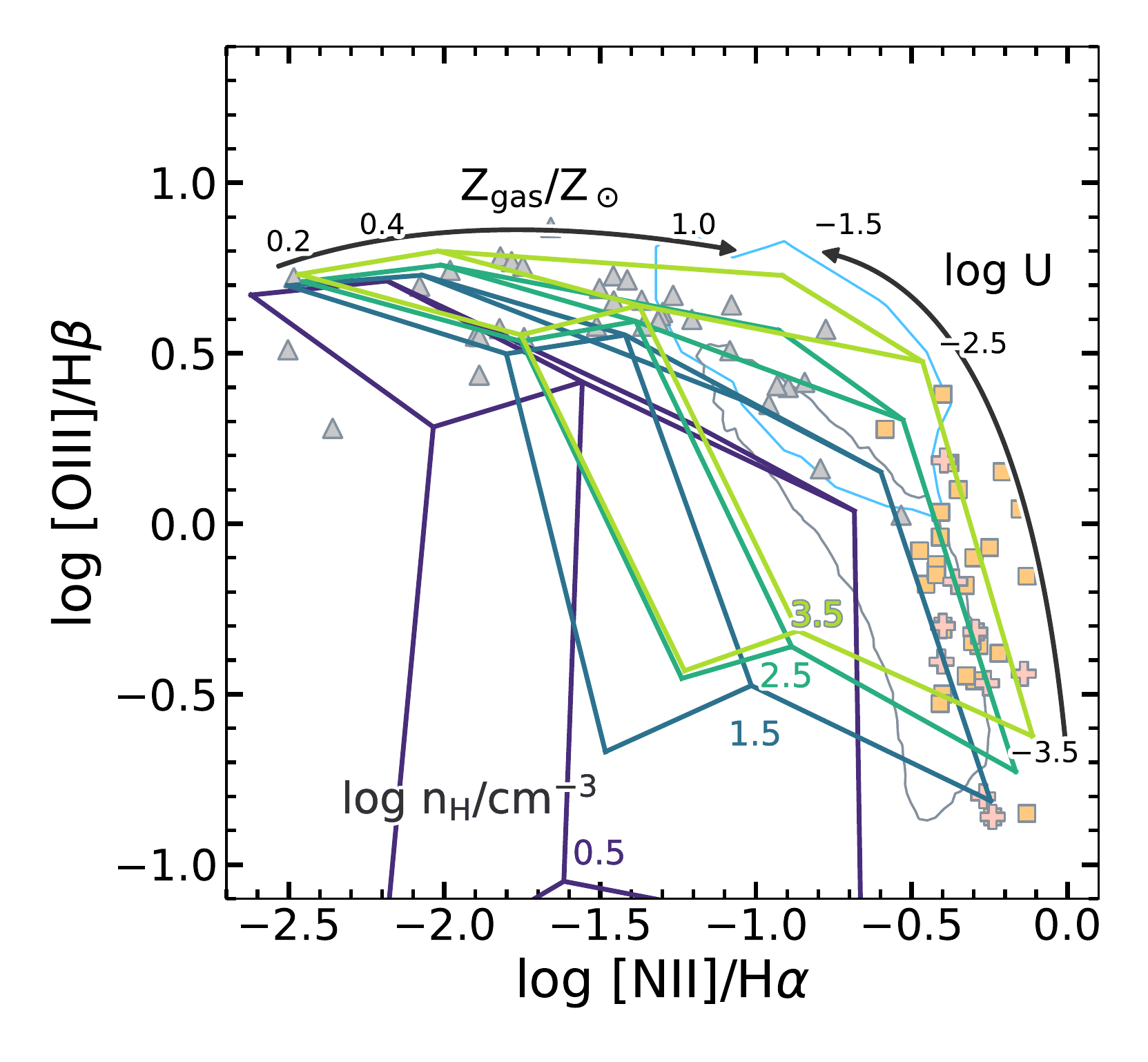}{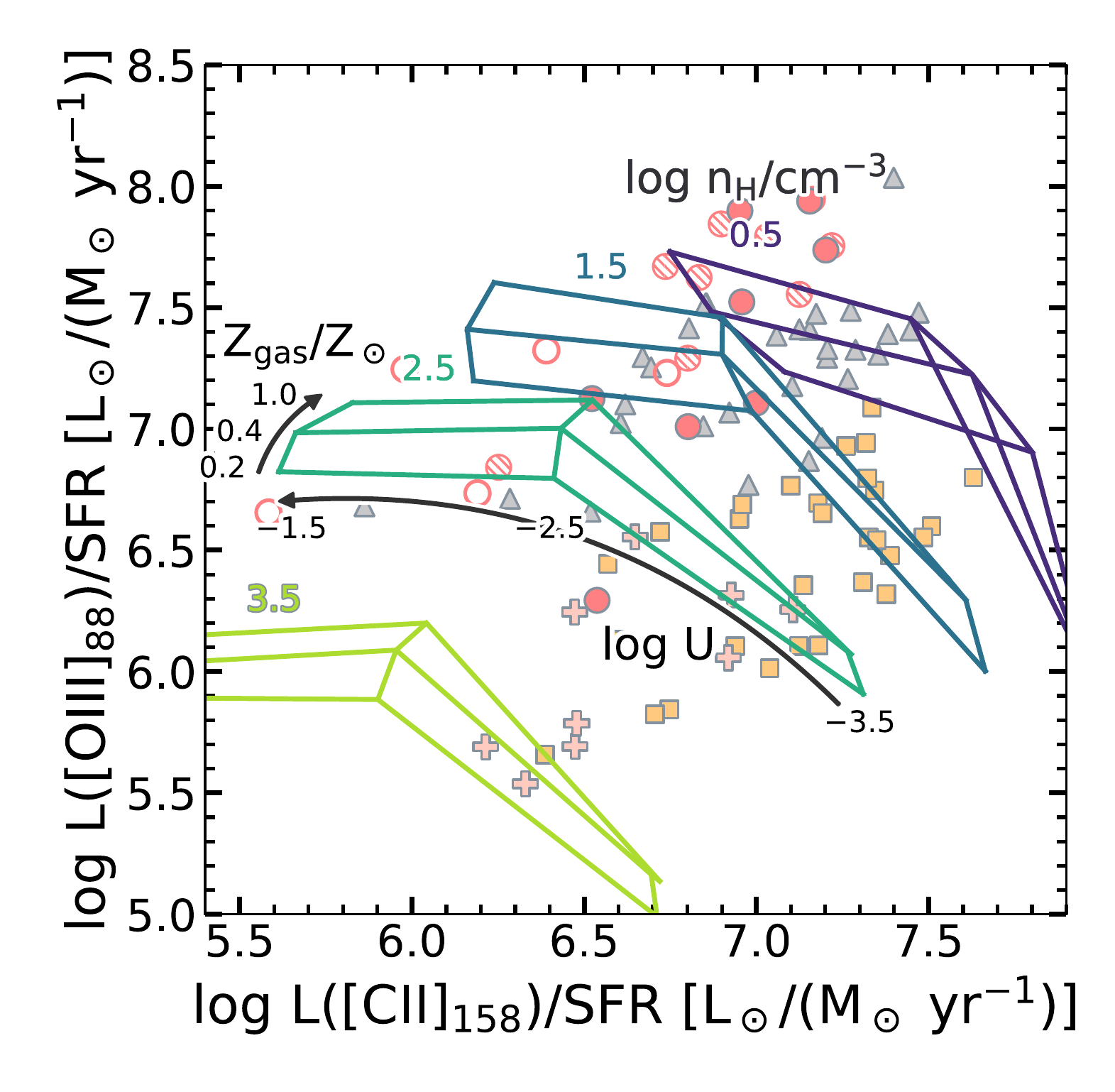}
    \caption{A part of fiducial photoionization model grids on the BPT (Left) and FIR (Right) diagrams.
      The input ionizing spectrum is the SB99 \( 300 \) Myr constant star-formation model.
      Each model grid indicates \(\log U = -1.5\), \(-2.5,\) and \(-3.5\) and \(\Zgas/\Zsun = 0.2\), \(0.4,\) and \(1.0\).
      The colors of the four grids indicate \(\log n_\text{H}/\cmmm = 0.5\), \(1.5\), \(2.5\), and \(3.5\) from purple (dark) to yellow (light).
      The symbols are the same as Figure \ref{fig:bptfir}, but we removed the symbols with gray edges, removed the error bars, and lightened colors.
      The gray and cyan contours in the left panel show the 95 and 68 percentile contours for the \(z\sim0\) and \(2\) galaxies, respectively.
      The model grids can explain most of the data points well with the three parameters, \(U\), \Zgas, and \(n_\text{H}\).
    }
  \label{fig:bptfir_grid}
\end{figure*}

\begin{deluxetable*}{@{\extracolsep{4pt}}lccccc}
    \tablecaption{Assumed parameters of the fiducial photoionization models.\label{tab:1}}
    \tablewidth{0pt}
    \tablecolumns{6}
    \tablehead{
      \colhead{} & \multicolumn{2}{c}{optFIR sample} & \multicolumn{2}{c}{optical sample} & \colhead{FIR sample} \\
      \cline{2-3} \cline{4-5} \cline{6-6}
      \colhead{Parameters} & \colhead{Dwarfs} & \colhead{U/LIRGs} & \colhead{\(z\sim0\) galaxies} & \colhead{\(z\sim2\) galaxies} & \colhead{\(z > 6\) galaxies}
    }
    \startdata
    Ionizing spectra & BPASS & SB99 & SB99 & BPASS & BPASS \\
    \ldots Star-formation history & \(300\) Myr cSF  & \(300\) Myr cSF & \(300\) Myr cSF & \(300\) Myr cSF & \(300\) Myr cSF \\
    \ldots \(Z_{*}/\Zgas\) & 1.0 & 1.0 & 1.0 & 0.2 & 0.2 \\
    Range of \(n_\text{H}/\cmmm\)  & \(0.5 < \log n_\text{H} < 2.5 \) & \(0.5 < \log n_\text{H} < 3.5 \) & \(0.5 < \log n_\text{H} < 2.5 \) & \(1.5 < \log n_\text{H} < 3.0 \) & \(1.5 < \log n_\text{H} < 3.0 \) \\
    Constraints on \(U\) and \Zgas & - & - & Equation \ref{eq:1} & Equation \ref{eq:11} & \(\Zgas/\Zsun < 1.0\) \\
    \cutinhead{Common parameters}
    Geometry & \multicolumn{5}{c}{Plane parallel under constant pressure} \\
    Abundances & \multicolumn{5}{c}{\HII\ region abundances (Helium: Equation \ref{eq:3}; Nitrogen: Equation \ref{eq:6})} \\
    Backgrounds & \multicolumn{5}{c}{Cosmic radio-to-X-ray background depending on redshift and cosmic-ray background} \\
    Magnetic field & \multicolumn{5}{c}{\(30\) \uG} \\
    Nebular parameter grids\tablenotemark{a} & \multicolumn{5}{c}{\( \log{U} = [-4.0, -0.5] \) (\( 0.5 \) intervals); \( \log{n_\text{H}/\mathrm{cm^{-3}}} = [0, 4.0] \) (\( 0.5 \) intervals); \(\Zgas/Z_{\odot} = 0.05, 0.2, 0.4, 1.0\), and \(2.0\)} \\
    Stopping criteria & \multicolumn{5}{c}{V-band dust extinction of 100 mag} \\
    \CPDR\tablenotemark{b} & \multicolumn{5}{c}{\(1.0\)} \\
    \enddata
    \tablenotetext{a}{When using the photoionization models, we interpolated the model grids into 20 steps to obtain finer grids.}
    \tablenotetext{b}{The PDR covering fraction is not an input parameter of \textsc{Cloudy} (see Section \ref{sec:param-depend-photo}).}
\end{deluxetable*}

We used \textsc{Cloudy} version 17.02 \citep{Ferland:1998a, Ferland.G:2017a} to construct photoionization models.
One of the major assumptions in our analysis is that a single nebular parameter set can reproduce the optical and FIR emission-line ratios simultaneously.
Therefore, we ignored contributions from the secondary component like diffuse ionized gas (DIG).

Following assumptions are adopted as fiducial models, which partly share the concepts with \citet{Nagao.T:2011a, Nagao.T:2012a}, \citet{Harikane.Y:2020b} and \citet{Sugahara.Y:2021a}.
The adopted assumptions are summarized in Table \ref{tab:1}.
A nebular structure is a plane-parallel geometry under the constant pressure.
Elemental and dust grain abundances are the \HII\ region abundances, but for helium and nitrogen.
A helium abundance reflects a combination of Big Bang nucleosynthesis and stellar yields, given by
\begin{equation}
\label{eq:3}
\text{He}/\text{H} = 0.0737 + 0.0293 \Zgas/\Zsun
\end{equation}
\citep{Groves.B:2004a}, where \Zgas\ is the gaseous metallicity and \Zsun\ is the solar metallicity of \(0.02\).
A nitrogen abundance would reflect a combination of primary and secondary nucleosynthesis as a function of metallicity \citep{Pilyugin.L:2012a, Andrews.B:2013a}.
Here, all over the redshift, \NO\ ratios keep constant at low \Zgas\ and then follow a local relation of extragalactic \HII\ regions \citep{Pilyugin.L:2014a}:
\begin{align}
\label{eq:6}
  \begin{array}{ll}
  \log{\NO} = -1.5 & (\mathcal{Z} \leq -0.59), \\
  \log{\NO} = 1.47 \mathcal{Z} - 0.656 & (\mathcal{Z} > -0.59),
  \end{array}
\end{align}
where \(\mathcal{Z} \equiv \log \Zgas/\Zsun\), the solar metallicity is assumed to be \(12 + \log(\OH) = 8.69\) \citep{Asplund.M:2009a}, and thus the criterion \(\mathcal{Z} = -0.59\) corresponds to \(12 + \log(\OH) = 8.1\).
Equation \ref{eq:6} gives \NO\ values comparable to those of the KBSS-MOSFIRE galaxies at \(z\sim2\) \citep{Strom.A:2018a} and that of B14-65666 at \(z = 7.15\) \citep{Sugahara.Y:2021a}.
A carbon abundance is as default in the fiducial models.
Our calculations extend to the PDR regions, from which a \CIIfir\ emission line mainly arises \citep{Russell.R:1980a, Tielens.A:1985a}.
Therefore, the fiducial models require parameters important for the PDR regions: the cosmic radio-to-X-ray background as a function of redshift \citep{Ostriker.J:1983a, Ikeuchi.S:1986a, Vedel.H:1994a}, ionization by cosmic rays background, the magnetic field of \(30\) \uG, and the PDR covering fraction (\CPDR) of unity.
As strengths of the cosmic rays background and magnetic fields are uncertain for high-redshift galaxies and even for local galaxies, a default value in \textsc{Cloudy} \citep{Indriolo.N:2007a} and a typical value in nearby spiral galaxies \citep{Beck.R:2015a} are applied, respectively.
The code stops the calculations at the \textit{V}-band dust extinction of 100 mag \citep{Abel.N:2005a}.

Input ionizing spectra varies according to galaxy populations.
We used two simple stellar population (SSP) models: \textsc{Starburst99} \citep[SB99;][]{Leitherer:1999a, Leitherer:2014a} with the non-rotating standard Geneva tracks and the Binary Population and Spectral Synthesis code \citep[BPASS,][]{Eldridge.J:2017a} version 2.2.1 \citep{Stanway.E:2018a}.
Both models aim to reproduce young starbursts, but the BPASS models exhibit harder ionizing spectra than the SB99 models owing to massive star binaries and thus produce higher \([\OIII]/\Hb\) and \(\LOiii/\SFR\) ratios than the SB99 models.
The BPASS models are applicable to high-redshift galaxies \citep[e.g.,][]{Steidel:2016}, which agrees with an expectation that young low-metallicity galaxies, typically seen at high-redshift, host many massive binaries \citep{Stanway.E:2020a}.
For the local U/LIRGs (optFIR sample) and \(z\sim0\) galaxies (optical sample), we applied a SB99 \(300\) Myr constant star-formation (cSF) model.
For the local dwarfs (optFIR sample), which would be analogs of high-redshift galaxies, we applied a BPASS \( 300 \) Myr cSF model.
These local dwarfs, U/LIRGs, and \(z\sim0\) galaxies are assumed to have stellar metallicity, \(Z_{*}\), equal to \Zgas.
On the other hand, FUV and optical spectra of galaxies at \(z\sim2\) indicate an importance of low \(Z_{*}/\Zgas\) ratios, which would reflect the abundance pattern of core-collapse supernovae yields, as well as massive star binaries \citep{Steidel:2016, Trainor.R:2016a}.
Accordingly, we used BPASS \( 300 \) Myr cSF model with \(Z_{*}/\Zgas = 0.2\) for the \(z\sim2\) galaxies (optical sample).
For the \( z > 6 \) galaxies (FIR sample), we assumed the identical ionizing spectra as for the \(z\sim2\) galaxies.
This assumption is supported by low \(Z_{*}/\Zgas\) ratios for galaxies at \(z > 2\) \citep{Cullen.F:2019a, Harikane.Y:2020a, Kashino.D:2022a}.
Section \ref{sec:implications-o3} will revisit our choices of the input ionizing spectra.

The fiducial models have three free nebular parameters: the ionization parameter \( U \), the hydrogen density \( n_\text{H} \), and the gas metallicity \Zgas.
The ionization parameter at the illuminated surface, \( U \), is a dimensionless parameter that expresses relative amount of ionizing photons to gas density, defined as \(U = \dot{n}_{\gamma}/n_\text{H} c\), where \( \dot{n}_{\gamma}  \) is the number flux of the ionizing photons and \( c \) is the speed of light.
\( U \) ranged from \( \log{U} = -4.0\) to \( -0.5  \) at \( 0.5 \) steps.
The hydrogen density at the illuminated surface ranged from \( \log{n_\text{H}/\mathrm{cm^{-3}}} = 0 \) to \( 4.0 \) at \( 0.5 \) steps.
Finally, the gas metallicity ranged \(\Zgas/Z_{\odot} = 0.05, 0.2, 0.4, 1.0\), and \(2.0\), where the solar metallicity value is \(Z_{\odot} = 0.02\).
For each nebular parameter set of (\(U\), \(n_\text{H}\), \(\Zgas\)), our models computed [\OIII]\lam5007, [\NII]\lam6548, \Ha, \OIIIfir, and \CIIfir\ line intensities relative to \Hb\ line.
As the FIR diagram requires \SFR\ in denominators of the two axes, we computed model line-to-\SFR\ ratios by converting \Hb\ line intensity to \SFR\ with Equation \ref{eq:10}.
Section \ref{sec:sfr-conversion} discusses the derivation of the conversion factor in detail.

Figure \ref{fig:bptfir_grid} illustrates a part of our fiducial model grids.
The illustrated model uses the SB99 \( 300 \) Myr cSF ionizing spectrum and shows grids at \(\log U = -1.5\), \(-2.5,\) and \(-3.5\); \(\Zgas/\Zsun = 0.2\), \(0.4,\) and \(1.0\); and \(\log n_\text{H}/\cmmm = 0.5\), \(1.5\), \(2.5\), and \(3.5\).
Our model grids can reproduce most observational data points in both of the diagrams.
While the model grids are relatively insensitive to \(n_\text{H}\) in the BPT diagram, a wide range of \(n_\text{H}\) is necessarily to explain the distributions in the FIR diagram.
This is because the FIR fine-structure lines are more sensitive to the electron (hydrogen) densities than the optical lines due to their lower critical densities.
Some previous studies fixed hydrogen densities at a typical values in their photoionization models \citep[e.g.,][]{Strom.A:2018a, Sanders.R:2020a}; this assumption would be reasonable for the BPT diagram, but not appropriate for the analysis of the FIR diagram.
The more detailed model dependencies on the BPT and FIR diagrams are discussed in \citet{Kewley.L:2013b} and \citet{Harikane.Y:2020b}, respectively.

\subsection{Modeling galaxy distributions}
\label{sec:model-galaxy-distr}
We search model solutions of the nebular parameters (\( U\), \(n_\text{H}\), and \(\Zgas\)) that can reproduce the distributions of galaxies in both of the diagrams.
As seen in Section \ref{sec:observational-data}, galaxy populations with similar characteristics are distributed in groups on the diagrams.
We aim to model not each galaxy but distributions of galaxy populations with the fiducial photoionization models.

In our modeling, the distributions of the galaxy populations are represented by regions defined as follows.
We defined the regions by hand for the optFIR (the local dwarfs and U/LIRGs) and FIR (the \(z > 6\) galaxies) samples.
These regions include most of the data points, but do not include some outliers.
The regions for the optical sample are defined from the number-density contours.
We used contours including 95\% and 68\% of the \(z\sim0\) and \(2\) galaxies, respectively.
We chose the relatively low-percentage contour for the \(z\sim2\) galaxies because the number of data points are still small and their measurement errors are large, compared with the \(z\sim0\) galaxies.
These defined regions are shown in Figure \ref{fig:bptfir} with the solid lines.

The searched nebular-parameter space is \( -4.0 < \log{U} < -0.5  \), \( 0.0 < \log{n_\text{H}/\cmmm} < 4.0\), and \( 0.1 < \Zgas/Z_{\sun} < 2.0 \).
The model grids of \(U\), \(n_\text{H}\), and \Zgas\ are interpolated into 20 steps to give finer grids.
The parameter space of each sample is additionally restricted according to the literature.
The hydrogen density is conservatively restricted to be \(0.5< \log{n_\text{H}} < 2.5\) for the local dwarfs \citep{Cormier.D:2019a, Spinoglio.L:2022a}, \( 0.5 < \log{n_\text{H}} < 3.5\) for the U/LIRGs \citep{Gracia-Carpio.J:2011a, Inami.H:2013a, Herrera-Camus.R:2016a, Zhao.Y:2016a}, \(0.5< \log{n_\text{H}} < 2.5\) for the \(z\sim0\) galaxies, and \( 1.5 < \log{n_\text{H}} < 3.0\) for the \(z\sim2\) galaxies \citep{Masters.D:2014a, Sanders.R:2016a}.
For the \(z > 6\) galaxies, we applied the same constraint as for the \(z\sim2\) galaxies because there are few measurements of \(n_\text{H}\) for the \(z > 6\) galaxies.
The ranges of \(U\) and \Zgas\ were restricted for the optical and FIR samples.
As shown in Section \ref{sec:pred-fir-diag}, the nebular parameters for the optical sample are poorly constrained only from the optical measurements on the BPT diagram.
To complement the lack of the observational constraints on the FIR diagram, we restricted \(U\) and \Zgas\ for the optical sample to follow \(U\)--\Zgas\ relations observed at \(z\lesssim2\) \citep[e.g.,][]{Perez-Montero.E:2014a}.
Whereas a local \(U\)--\Zgas\ relation seems to hold at \(z\sim2\) at \(\Zgas/Z_{\odot} < 1/3\) \citep[][]{Sanders.R:2020a}, \(U\) values at \(z\gtrsim1\) seem to be \(0.2\nn0.5\) dex higher than local at \(\Zgas/Z_{\odot} \sim 1\) \citep{Nakajima:2014, Kashino.D:2017a}.
Therefore, we analytically expressed a \(U\)--\Zgas\ relation of \citet{Perez-Montero.E:2014a} and employed it for the \(z\sim0\) galaxies,
\begin{equation}
  \label{eq:1}
  -0.83 \mathcal{Z} - 3.6 < \log U < -1.67 \mathcal{Z} - 3.0\ \ (z\sim0).
\end{equation}
We then modified the relation for the \(z \sim 2 \) galaxies,
\begin{equation}
\label{eq:11}
  -0.83 \mathcal{Z} - 3.6 < \log U < -1.17 \mathcal{Z} - 2.7\ \ (z\sim2),
\end{equation}
where the upper limit is \(0.3\) dex higher than Equation \ref{eq:1} at \(\mathcal{Z} = 0\).
Even if the increase of the upper limit becomes \(0.2\nn0.5\) dex, our results do not change.
We used the same lower limit as Equation \ref{eq:1} because the increase in the lower limit makes it difficult to model the \(z\sim2\) galaxies exhibiting relatively low \([\OIII]/\Hb\) ratios.
These modifications are consistent with a recent observational study of the \(U\nn\Zgas\) relation at \(z\sim1\nn2\) \citep{Papovich.C:2022a}.
For the FIR sample, the \( z > 6 \) galaxies were assumed to have sub-solar metallicities, because high-redshift galaxies likely have low metallicities according to the redshift evolution of the mass--metallicity relation \citep[e.g.,][]{Zahid:2013b}.

\section{Results}
\label{sec:results}

\subsection{Nebular parameters for the optFIR sample}
\label{sec:relat-betw-bpt}

\begin{figure*}[t]
    \epsscale{1.2}
    \plotone{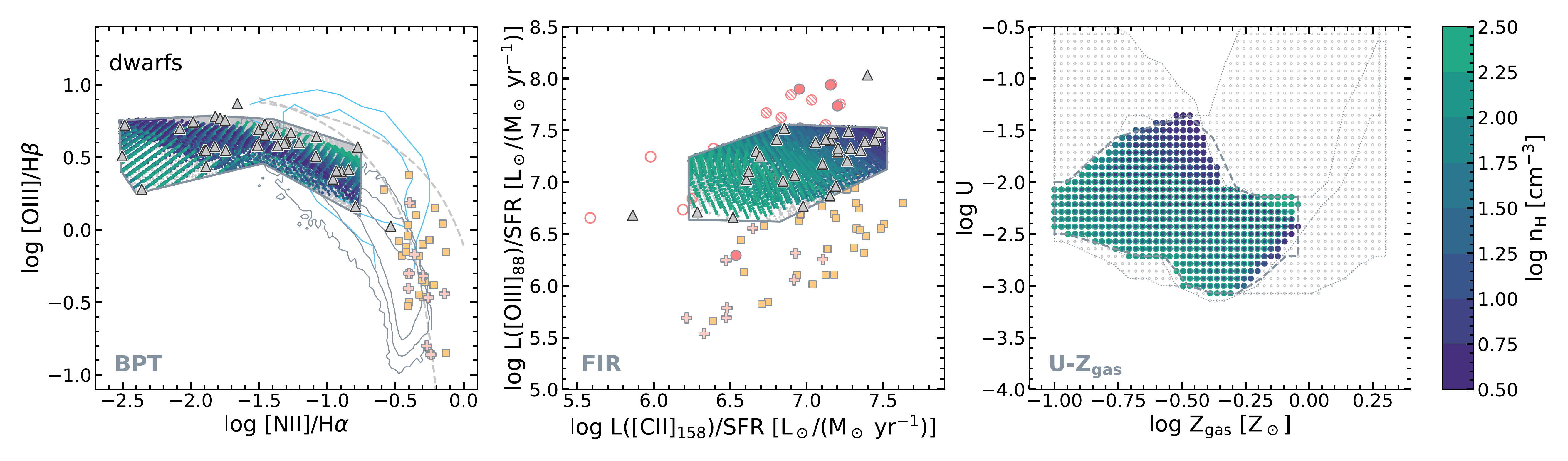}
    \plotone{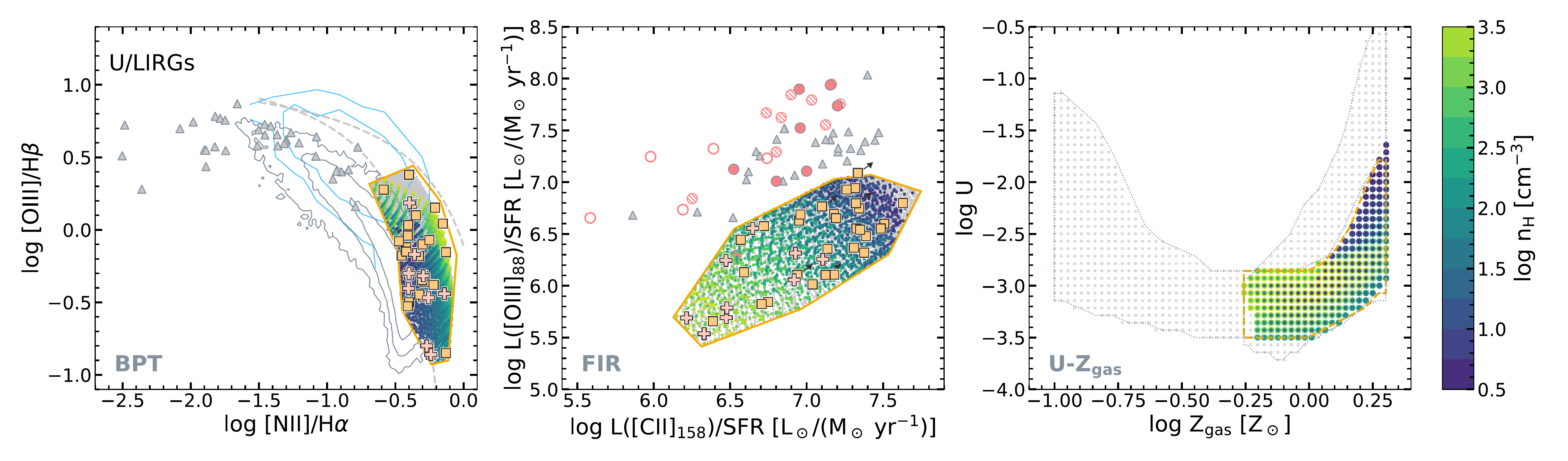}
    \caption{Solutions of the photoionization modeling for the optFIR sample: the dwarfs (Top) and the U/LIRGs (Bottom).
      We searched the nebular parameters that satisfy the regions in the BPT (Left) and FIR (Middle) diagrams and plot the obtained parameters in the \(U\nn \Zgas\) diagram (Right).
      \textit{Left and Middle}: The obtained solutions are plotted with being color-coded by \(n_\text{H}\).
      The symbols and contours are the same as Figure \ref{fig:bptfir_grid}, but the symbols of the target galaxies are highlighted.
      \textit{Right}: The colored circles show the obtained solutions, of which outer and inner colors show the maximum and minimum \(n_\text{H}\), respectively, at a given \(U\) and \(\Zgas\).
      The gray data points are model solutions that satisfy one of the regions in the BPT or FIR diagrams, but not the other.
      The obtained nebular-parameter regions are illustrated with the dashed polygons.
      We note that model solutions for the dwarfs around \((\log{\Zgas/\Zsun}, \log{U})\sim(0.2, -1.3)\) are ignored because these are far from the main solutions
    }
  \label{fig:result-optfir}
\end{figure*}

Figure \ref{fig:result-optfir} shows the obtained nebular-parameter solutions for the optFIR sample that satisfy the regions in the BPT (left) and FIR (middle) diagrams.
The right panels show the obtained nebular parameters (\(U\), \Zgas, \(n_\text{H}\)).
The optFIR sample helps to check whether our photoionization model reproduces proper nebular parameters of local galaxies from the BPT and FIR diagrams.

The top panels show the results for the local dwarfs.
The dwarfs exhibit low [\NII]\(/\)\Ha\ and high [\OIII]\(/\)\Hb\ in the BPT diagram and high \(\LOiii/\SFR\)\ in the FIR diagram, suggesting high ionization and low metallicity environments.
In fact, our photoionization model gives \(-3.0 < \log{U} < -1.5 \) and sub-solar metallicity (\( \Zgas/\Zsun < 1.0 \)), as shown in the right panel.
The range of the obtained nebular parameters is consistent with those measured for the DGS galaxies \citep{Cormier.D:2015a, Cormier.D:2019a}, although some of the dwarfs are measured to have very low metallicities of \(\Zgas/\Zsun < 0.1 \) \citep{Madden.S:2013a}.
We note that there are model solutions around \((\log{\Zgas/\Zsun}, \log{U})\sim(0.2, -1.3)\), but we ignore these solutions because these are far from the main solutions.

The bottom panels show the results for the local U/LIRGs.
The U/LIRGs exhibit high [\NII]\(/\)\Ha\ and low [\OIII]\(/\)\Hb\ in the BPT diagrams and low \(\LOiii/\SFR\)\ in the FIR diagrams, indicating that these galaxies have nebular properties different from the dwarfs.
In the right panel, the positive correlation between \(U\) and \(\Zgas\) is tight despite a wide range of \( n_\text{H} \); \( \log U \) increases from \(-3.5\) to \(-2.0\) as \( \log \Zgas/\Zsun \) increases from \(-0.25\) to \(0.3\).
This positive \(U\)--\Zgas\ correlation is opposite to the negative correlations of the local dwarfs, \HII\ regions, and galaxies at \(z\sim0\) \citep{Perez-Montero.E:2014a, Kashino.D:2019a}, but these negative \(U\nn\Zgas\) relations smoothly connected to the distributions of the U/LIRGs.
Specifically, the nebular parameter space for the U/LIRGs do not share with that for the dwarfs.
Our results of \( U \) and \( \Zgas \) are roughly consistent with previous results using MIR and FIR lines \citep[\(-3.2 \lesssim \log{U} \lesssim - 2.0 \) and \(0.7 \lesssim \Zgas/Z_{\odot} \lesssim 2.0 \);][]{Gracia-Carpio.J:2011a, Inami.H:2013a, Diaz-Santos.T:2017a, Pereira-Santaella.M:2017a}.
However, there is a caution that emission from the U/LIRGs may be still contaminated by AGNs.
As our models assume only ionizing spectrum of star formation, we would be careful to interpret the results for the U/LIRGs.

Comparisons with previous studies support that our fiducial photoionization models can reproduce the galaxy nebular parameters from the BPT and FIR diagrams.
In the next sections, we apply these models to the optical and FIR samples to predict galaxy distributions on the FIR and BPT diagrams, respectively.

\subsection{FIR diagram at \( z\sim0\nn2 \) for the optical sample}
\label{sec:pred-fir-diag}

\begin{figure*}[t]
    \epsscale{1.2}
    \plotone{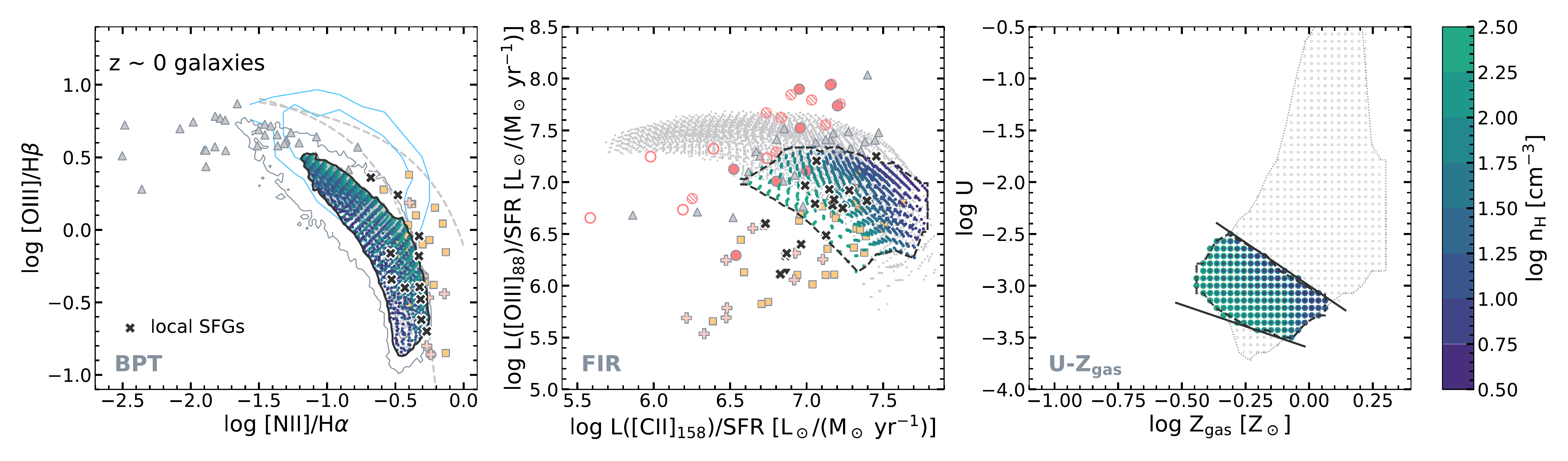}
    \plotone{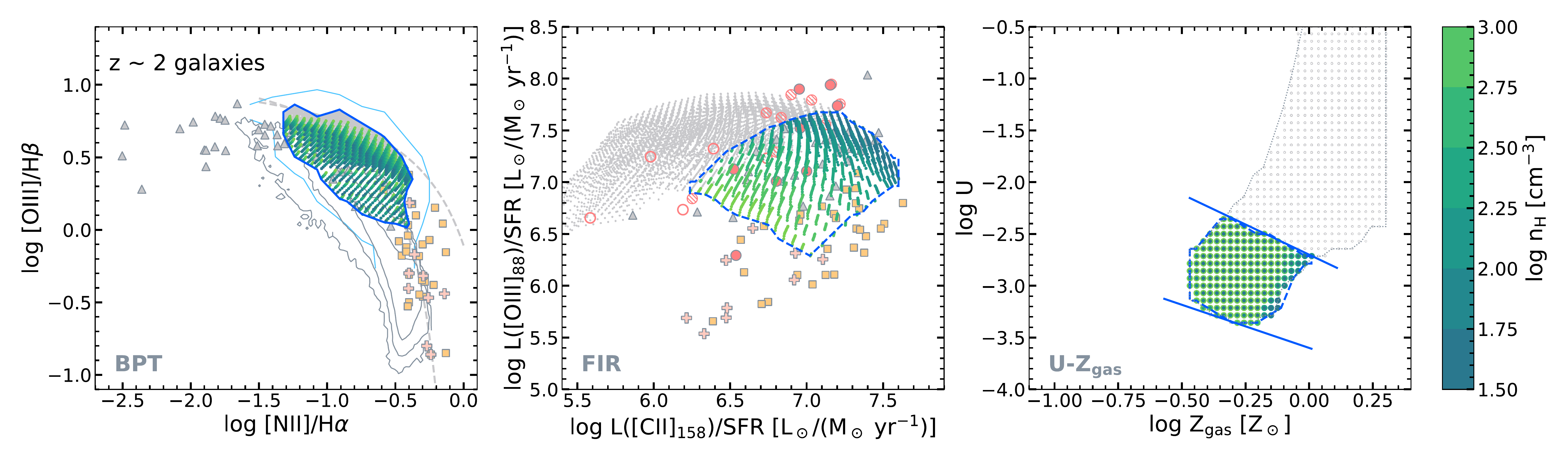}
    \caption{Solutions of the photoionization modeling for the optical sample: the \(z\sim0\) galaxies (Top) and the \(z\sim2\) galaxies (Bottom).
      We searched the nebular parameters that satisfy the regions in the BPT diagrams (Left) and the constraints on the nebular parameters and plot the obtained solutions in the FIR (Middle) and \(U\nn \Zgas\) (Right) diagrams.
      The cross symbols in the top panels indicate the \textit{ISO} galaxies \citep{Brauher.J:2008a}.
      \textit{Middle and Right}: The gray data points show the model parameters that satisfy the regions in the BPT diagram but do not the constraints on the nebular parameters (solid lines in the right panels).
      The obtained regions are illustrated with the dashed polygons.
      Other symbols are the same as Figure \ref{fig:result-optfir}.
    }
  \label{fig:result-opt}
\end{figure*}

\begin{figure*}[t]
    \epsscale{1.2}
    \plotone{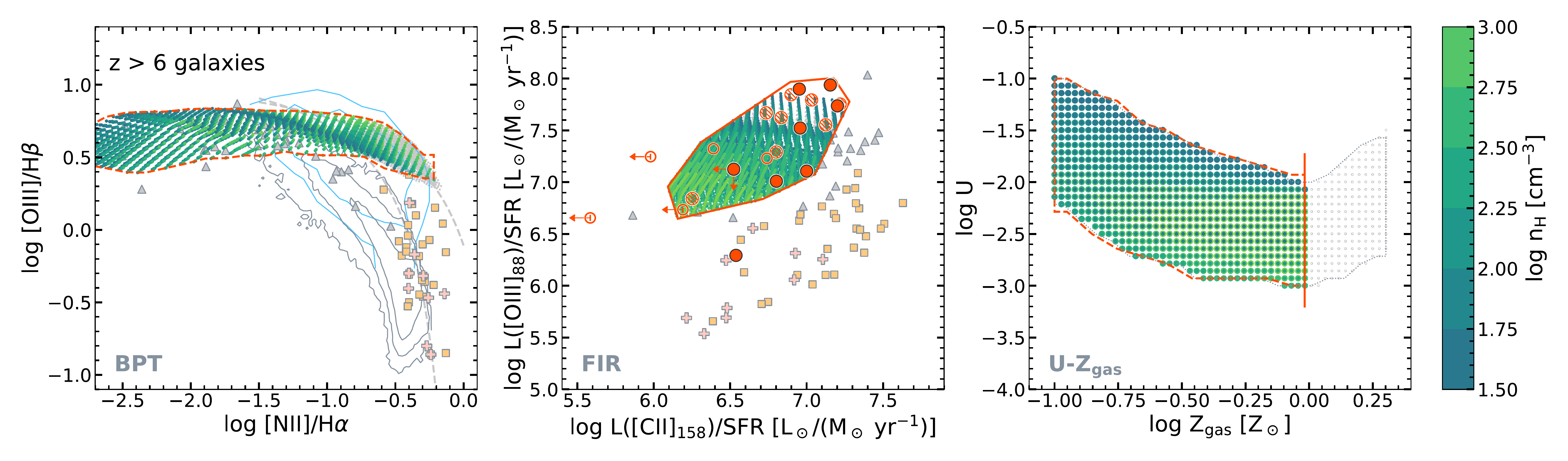}
    \caption{Solutions of the photoionization modeling for the \(z > 6\) galaxies (the FIR sample).
      We searched the nebular parameters that satisfy the regions in the FIR diagram (Middle) with \(Z < 1.0 Z_{\sun}\) and plot the obtained solutions in the BPT (Left) and \(U\nn \Zgas\) (Right) diagrams.
      \textit{Left and Right}: The gray data points show model parameters that satisfy the region in the FIR diagram, but at \(\Zgas/\Zo \geq 1\).
      The obtained regions are illustrated with the dashed polygons.
      The symbols are the same as Figure \ref{fig:result-optfir}, but the symbols of the \(z > 6\) galaxies are highlighted.
    }
  \label{fig:result-fir}
\end{figure*}

The top panels of Figure \ref{fig:result-opt} show the results for the \(z\sim0\) galaxies.
We searched the nebular parameters that satisfy the region in the BPT diagram (left panel) and the restrictions of \(U\) and \Zgas\ (right panel).
The range of the obtained nebular parameters are \(-3.5 < \log U < -2.5 \) and \(-0.45 < \log Z/Z_{\sun} < 0.1\).
The middle panel shows a distribution of the \(z\sim0\) galaxies in the FIR diagram predicted from the obtained nebular parameters.
To check the accuracy of our predictions, we plotted local star-forming galaxies observed with \textit{ISO} \citep{Brauher.J:2008a} with the cross symbols.
Their \SFR\ was taken from \citet{De-Looze.I:2014a}; their optical measurements were taken from the same literature as for the optFIR sample; and AGNs, U/LIRGs, and resolved objects were not included.
The distribution of the \textit{ISO} galaxies in the FIR diagram agrees well with the prediction for the \(z\sim0\) galaxies, while four \textit{ISO} galaxies located on the region of the local U/LIRGs.
This good agreement supports results and predictions of our photoionization modeling.

The bottom panels show the results for the \(z\sim2\) galaxies.
The obtained \( U \) and \Zgas\ values are similar to those for the \(z\sim0\) galaxies, but slightly lower \Zgas\ values are dominant.
The predicted distribution in the FIR diagram shows a weak positive correlation between \(\LOiii/\SFR\) and \(\LCii/\SFR\), similar to the local dwarfs and U/LIRGs.
We note that, if there exist galaxies at \(z\sim2\) having high hydrogen density of \(\log{n_\text{H}/\cmmm} > 3.0\) \citep{Lehnert.M:2009a, Hainline.K:2009a, Bian.F:2010a}, they would be located at lower \(\LOiii/\SFR\) and \(\LCii/\SFR\) than the predicted distributions.
Compared with the \(z\sim0\) galaxies, the \(z\sim2\) galaxies are predicted to exhibit higher \(\OIIIfir/\CIIfir\).

For both of the \(z\sim0\) and \(2\) galaxies, the gray points in the \(U\)--\Zgas\ diagrams (right panels) show the non-selected nebular parameters that explain the regions in the BPT diagram but do not satisfy the restrictions of the \(U\) and \Zgas.
These non-selected nebular parameters demonstrate that it is difficult to constrain the nebular parameters only from the BPT diagram, due to the parameter degeneracy between (low \(U\), low \Zgas, high \(n_\text{H}\)) and (high \(U\), high \Zgas, low \(n_\text{H}\)) (see Figure \ref{fig:bptfir_grid}).

One may concern why almost the same \(U\) and \Zgas\ values for the \(z\sim0\) and \(2\) galaxies can reproduce the different distributions in the BPT diagrams.
These differences actually originate from the different assumptions in our photoionization models: the input ionizing spectra and hydrogen density.
In Section \ref{sec:photo-ioniz-model}, we assumed ionizing spectra of BPASS models with \(Z_{*}/\Zgas = 0.2\) for the \(z\sim2\) galaxies.
These massive star binaries and low stellar metallicities make the spectrum harder than the ionizing spectra of SB99 models, which was used for the \(z\sim0\) galaxies.
The hard spectra and high hydrogen (i.e., electron) densities increase both \([\OIII]/\Hb\) and \([\NII]/\Ha\) in the BPT diagram \citep{Kewley.L:2013b}.
In our photoionization models, thus, we explain the redshift evolution from \( z\sim0 \) to \(2\) with assumed harder ionizing spectra and higher hydrogen density at \( z\sim2 \), as well as slightly higher \(U\) and lower \Zgas.

\subsection{Nebular parameters and BPT diagram at \( z>6\) for the FIR sample}
\label{sec:pred-nebul-param}

\begin{figure}[t]
    \epsscale{1.2}
    \plotone{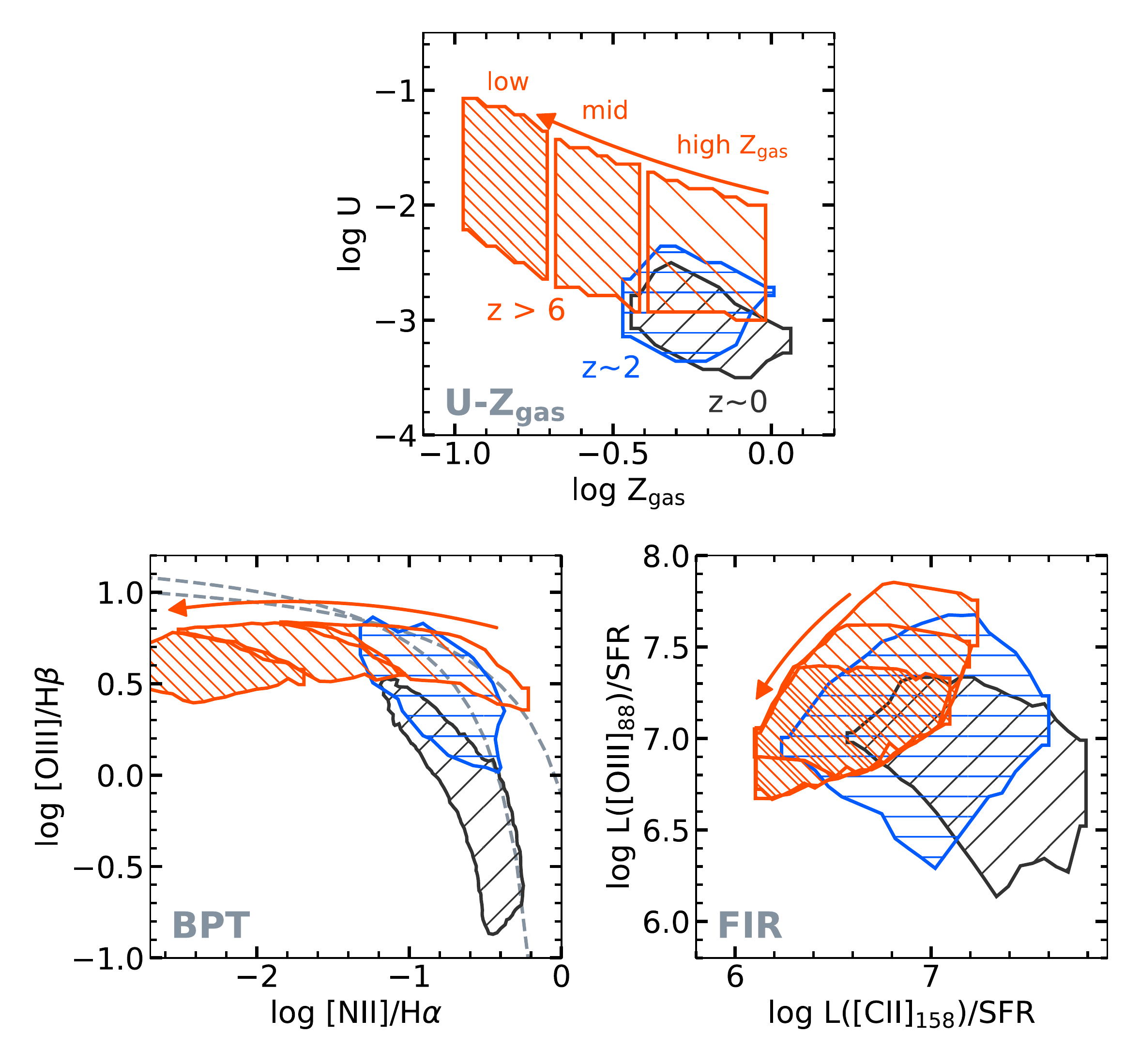}
    \caption{Metallicity dependence of the regions of the \(z > 6\) galaxies on the \(U\nn\Zgas\) (Top), BPT (Bottom left), and FIR (Bottom right) diagrams.
      The red hashed regions of the \(z > 6\) galaxies are divided into three regions: \(-1.0 \le \mathcal{Z} < -0.7\) (low), \(-0.7 \le \mathcal{Z} < -0.4\) (mid), \(-0.4 \le \mathcal{Z} < 0.0\) (high), where \(\mathcal{Z} \equiv \log \Zgas/\Zsun\).
      We note that, on the FIR diagram, higher-\Zgas\ regions overlap with lower-\Zgas\ regions.
      The black and blue regions of the \(z \sim 0\) and \(2\) galaxies, respectively, are drawn for comparison.
    }
  \label{fig:result-metal}
\end{figure}

\begin{figure*}[t]
    \epsscale{0.38}
    \plotone{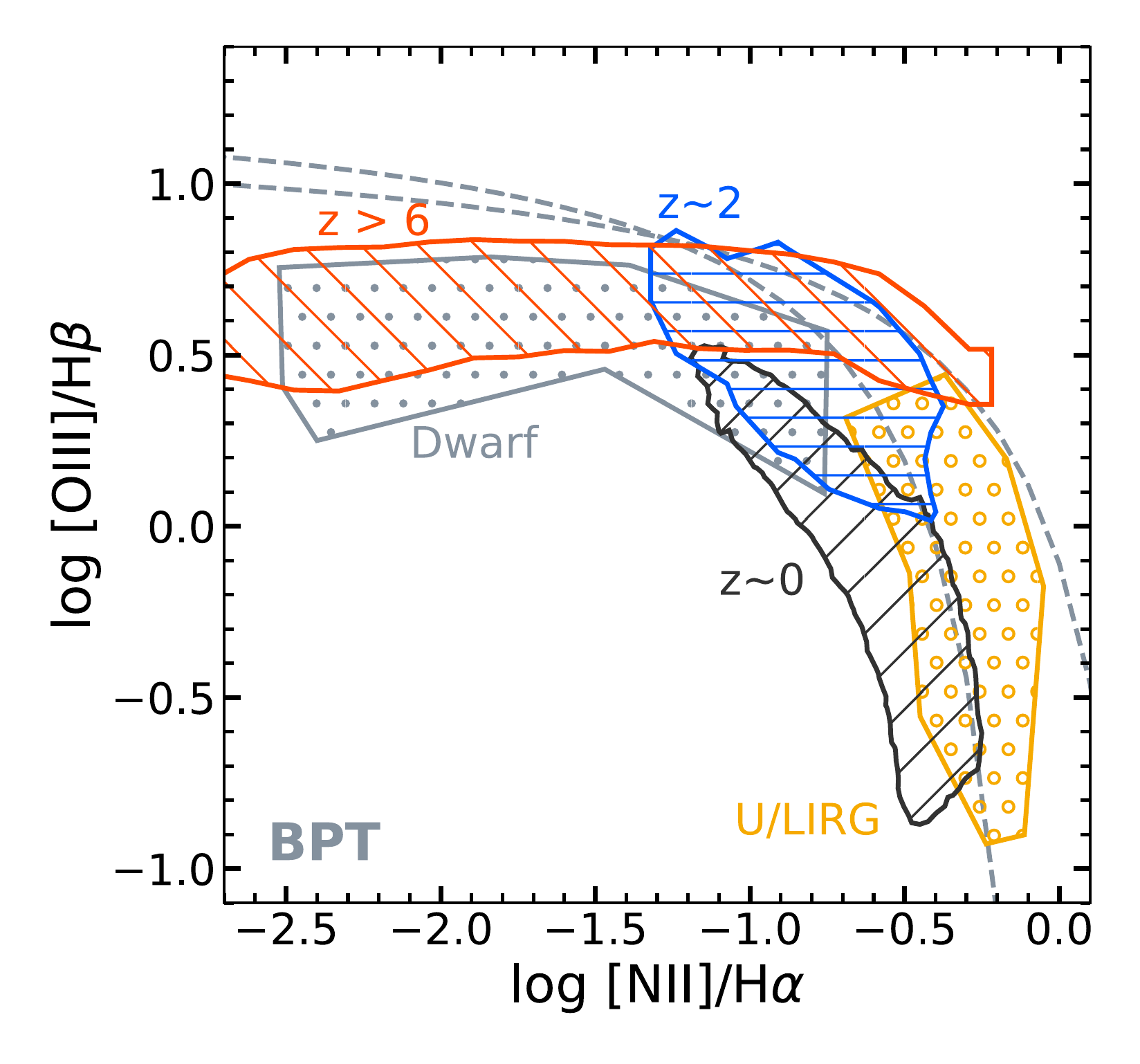}
    \plotone{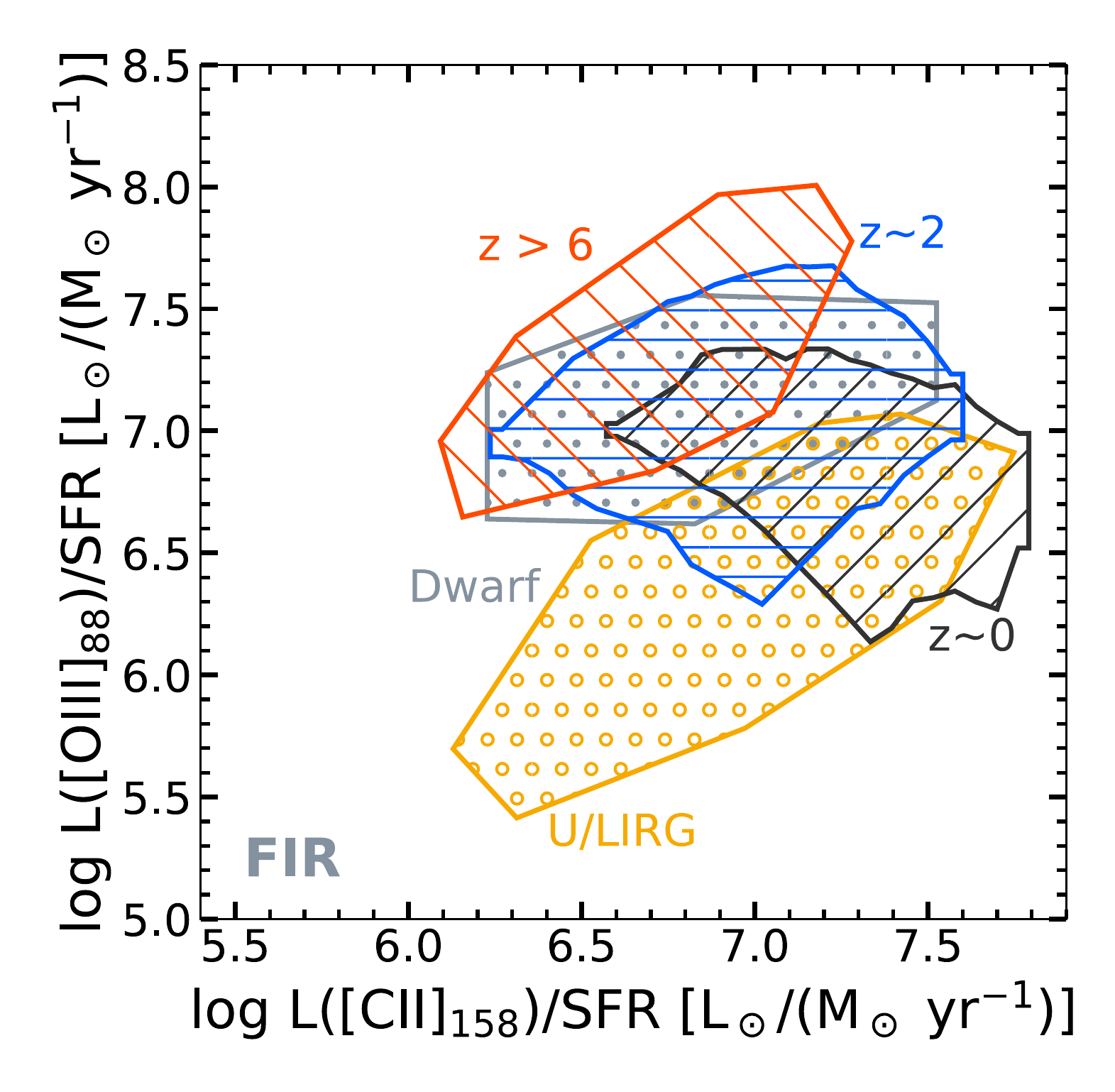}
    \plotone{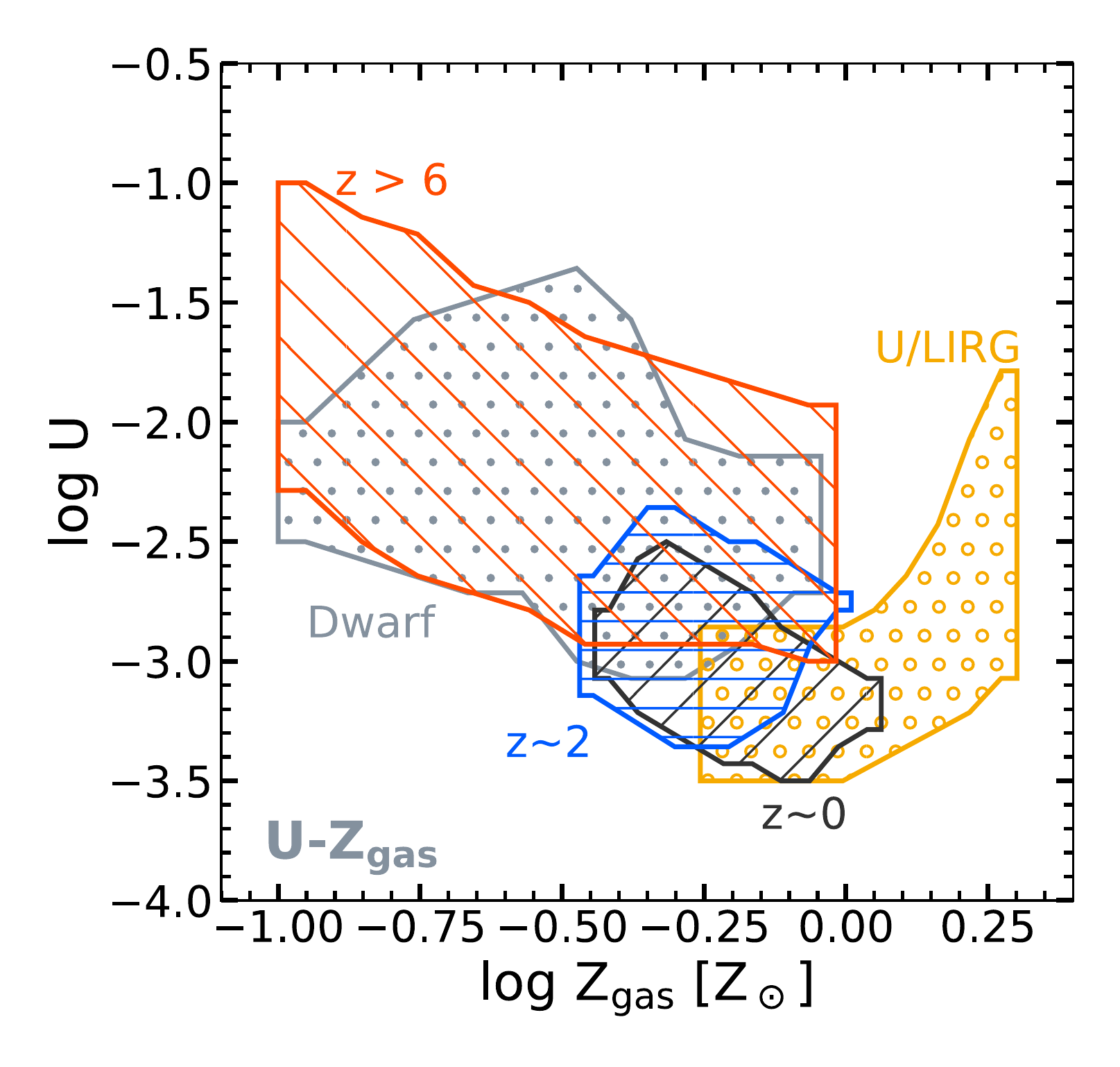}
    \caption{Graphical summary of the regions where the galaxy populations are distributed on the BPT (Left), FIR (Middle), and \(U\nn \Zgas\) (Right) diagrams.
      The regions show the distributions of the dwarfs (gray), U/LIRGs (orange), \(z\sim0\) (black), \(z\sim2\) (blue), and \(z > 6\) (red) galaxies.
      These regions are either drawn by hand on the basis of the galaxy distributions or obtained from the fiducial photoionization modeling.
      The gray dashed lines depict the criteria between star-forming galaxies and AGNs \citep{Kauffmann.G:2003a, Kewley.L:2001b}.
    }
  \label{fig:result-summary}
\end{figure*}

Figure \ref{fig:result-fir} shows the results for the FIR sample (i.e., the \(z > 6\) galaxies).
We searched the nebular parameters that satisfy the region in the FIR diagram for the \(z > 6\) galaxies.

The right panel of Figure \ref{fig:result-fir} shows that \( U \) and \Zgas\ have a negative correlation similar to those seen in galaxies at \( z\sim0 \) \citep[e.g.,][]{Perez-Montero.E:2014a, Kashino.D:2019a}.
Interestingly, this negative correlation is obtained from only the region on the FIR diagram.
The ionization parameter at low \Zgas\ is comparable to those of Lyman-alpha emitters at \(z\sim 2\) \citep{Nakajima:2014, Trainor.R:2016a, Kojima.T:2017a}.
The metallicity spans all the allowed range of \(-1.0 < \log{\Zgas/Z_{\odot}} < 0.0 \), implying that our analysis cannot find characteristic \Zgas\ for the \(z > 6\) galaxies.
To clarify the metallicity dependence of the \(z > 6\) galaxies, we divided the region into the three \Zgas\ range as illustrated in Figure \ref{fig:result-metal}.
The FIR diagram (bottom right panel) shows that an allowed range of \Zgas\ depends on \(\LOiii/\SFR\); namely, \Zgas\ is constrained to be high for galaxies with high \(\LOiii/\SFR\) whereas not constrained for galaxies with low \(\LOiii/\SFR\).
This dependence of the allowed \Zgas\ range is consistent with predictions of the analytical models in \citet{Yang.S:2020a}.
As proposed in \citet{Moriwaki.K:2018a}, \citet{Jones.T:2020a}, and \citet{Yang.S:2020a}, [\OIII] 52 \um\ and [\OIII] \lam5007 lines are important to resolve a degeneracy in \( U \), \( n_\text{H} \), and \Zgas\ at \( z > 6 \).

The predicted distribution of the \( z > 6\) galaxies in the BPT diagram is shown in the left panel of Figure \ref{fig:result-fir}.
The predicted \(\log{[\OIII]/\Hb}\) is almost constant at \( 0.5\nn0.8 \) whereas the predicted \(\log{[\NII]/\Ha}\) varies from \( < -2.5 \) to \( -0.3 \).
This predicted flat distribution is different from those of other low-redshift galaxy populations that follow the curved locus of star-forming galaxies (i.e., \([\OIII]/\Hb\) decreasing with \([\NII]/\Ha\)).
Upcoming NIR observations with the \textit{James Webb Space Telescope} (\textit{JWST}) will test our predictions for the BPT diagram.
Moreover, \([\NII]/\Ha\) significantly depends on \Zgas\ as shown in Figure \ref{fig:result-metal}.
This dependence suggests a possibility that the \([\NII]/\Ha\) ratio is still a metallicity tracer in the \textit{JWST} era \citep[N2 index;][]{Pettini.M:2004a, Nagao.T:2006c}, although the different nebular parameters at \(z > 6\) may cause systematics in metallicity measurements \citep{Bian.F:2018a}.

We would like to stress that our results are consistent with conclusions of \citet{Harikane.Y:2020b}, who used similar photoionization models.
They concluded that high \(U\), low \CPDR, or both are necessarily to explain the distributions in the FIR diagram for galaxies at \(z > 6\), especially for galaxies with high \(\OIIIfir/\CIIfir\) ratios.
Our results similarly predict high ionization parameters of \(\log U \simeq -2.5\) to \(-1.5\) for the \(z > 6\) galaxies at \(\CPDR = 1.0\).
A major difference is that the region for the photoionization modeling in our work does not include data points with the highest \(\OIIIfir/\CIIfir\) ratios, SXDF-NB1006-2 and MACS1148-JD1 (open red circles), because these measurements are based on \SFRSED.
Due to this difference, higher ionization parameters like \(\log U = -1\) are unnecessary in our modeling.

\subsection{Bridging galaxy populations}
\label{sec:bridge-galpop}

Figure \ref{fig:result-summary} summarizes the regions of the galaxy populations, observed and predicted from the modeling.
At \(z\sim0\), in all the panels, the distributions of the galaxy populations continuously shift from the local U/LIRGs through the normal \(z\sim0\) galaxies to the local dwarfs.
Although this trend has been already observed in the BPT diagram, the predicted distributions of the \(z\sim0\) galaxies bridge a bi-modality between the dwarfs and U/LIRGs in the FIR and \(U\nn\Zgas\) diagrams.
The gas metallicity monotonically increases from the dwarfs to the U/LIRGs and the ionization parameter gradually changes along with metallicity.
Combining the dwarfs and \(z\sim0\) galaxies (that follow Equation \ref{eq:1}) presents a negative \(U\nn\Zgas\) correlation, which supports that the nebular parameters for the dwarfs, obtained from our photoionization modeling, agree with the local \(U\nn\Zgas\) relation \citep[][]{Perez-Montero.E:2014a}.
In contrast, the U/LIRGs show a positive correlation that smoothly connects with the \(z\sim0\) galaxies at \(\log \Zgas/\Zsun \sim 0.1\).

The \(z\sim0\), \(2\), and \( > 6\) galaxies show the redshift evolution of the galaxy distributions in all the diagrams.
In the BPT diagram, the \( z\sim2 \) galaxies exhibit higher \([\OIII]/\Hb\) ratios than the \( z\sim0 \) galaxies, and the \( z > 6 \) galaxies will exhibit higher \([\OIII]/\Hb\) ratios than the \(z\sim2\) galaxies on average.
The \( z > 6 \) galaxies exhibit lower \([\NII]/\Ha\) ratios than the \(z\sim2\) galaxies; however, as shown in Figure \ref{fig:result-metal}, \([\NII]/\Ha\) ratios strongly depend on \Zgas\ and similar \Zgas\ (\(\log{\Zgas/\Zo} = -0.6\) to \(0\)) will predict similar \([\NII]/\Ha\) ratios at all the redshifts.
Thus, higher-redshift galaxies exhibit higher \([\OIII]/\Hb\) ratios at given \Zgas\ (or \([\NII]/\Ha\)).
In the FIR diagram, higher-redshift galaxies show higher \(\OIIIfir/\CIIfir\) ratios on average, which is consistent with previous findings at \(z\sim0\) and \( > 6\) \citep{Inoue.A:2016a, Hashimoto.T:2019a, Harikane.Y:2020b}.
The evolution in the BPT and FIR diagrams would reflect the redshift evolution of the nebular parameters and the hardness of ionizing spectra.
In the \(U\nn\Zgas\) diagram, the \(z > 6\) galaxies exhibit \(\sim\!\!0.5\) dex (up to \(1\) dex) higher \(U\) than the \(z\sim0\) and \(2\) galaxies at a given \Zgas, in good agreement with the results of \citet{Harikane.Y:2020b}.
This higher \(U\) and assumed harder ionizing spectra at higher redshift give rise to the higher \([\OIII]/\Hb\) and \(\OIIIfir/\CIIfir\) ratios in our photoionization models.

The stellar mass is important in discussions of the redshift evolution.
The average stellar mass of the \(z > 6\) galaxies may be several times less than those of the \(z\sim0\) and \(2\) galaxies (Section \ref{sec:observational-data}), but an increase in \(U\) caused by the low stellar mass would be less than \(0.25\) dex at given \Zgas\ and \(n_\text{H}\) according to local scaling relations \citep{Kashino.D:2019a}.
Therefore, the predicted increases in \(U\) for the \(z > 6\) galaxies are larger than expected from the difference of the stellar mass.

Specifically, the \(z > 6\) galaxies and local dwarfs share a large part of their distributions on the diagrams.
This similarity may support an argument that, regarding ISM properties, nearby dwarf galaxies are local analogues of high-redshift galaxies, despite differences of their masses.

\section{Discussion}
\label{sec:discussion}

\subsection{Implications of high \([\OIII]/\Hb\) and \(\LOiii/\SFR\)}
\label{sec:implications-o3}

In the fiducial photoionization models, we have assumed the ionizing spectra depending on the galaxy populations (Section \ref{sec:photo-ioniz-model}).
This section discusses the necessity of the input ionizing spectra to represent line ratios, especially for \([\OIII]/\Hb\) and \(\LOiii/\SFR\).

The ionizing spectra we assumed for the local dwarfs and \(z\sim2\) galaxies are the BPASS \(300\) Myr cSF models with \(Z_{*}/\Zgas = 1\) and \(0.2\), respectively; these hard spectra help to reproduce high \([\OIII]/\Hb\) ratios observed in these galaxy populations.
Both of the galaxy populations include galaxies with \(\log{[\OIII]/\Hb} > 0.7\) in their regions on the BPT diagram.
The SB99 cSF models are often used for modeling local galaxies \citep[e.g.,][]{Inami.H:2013a}, including the DGS galaxies \citep{Cormier.D:2019a}, and they are actually suitable to explain low \([\OIII]/\Hb\) ratios in the \(z\sim0\) galaxies and U/LIRGs in this study.
However, the SB99 models are unable to reproduce the high \([\OIII]/\Hb\) ratios in the local dwarfs because the ionization parameters are constrained by the region on the FIR diagram.
For this reason, the local dwarfs require harder input ionizing spectra than the SB99 models and prefer the BPASS models.
The requirement becomes more stringent for the \(z\sim2\) galaxies, which exhibit higher \([\NII]/\Ha\) ratios than the local dwarfs.
The BPASS cSF models at \(Z_{*}/\Zgas = 1\) can explain only a half of the region of the \(z\sim2\) galaxies on the BPT diagram with satisfying the \(U\nn\Zgas\) relation.
To explain the remaining galaxies having high \([\OIII]/\Hb\), harder ionizing spectra due to low \(Z_{*}/\Zgas\) are necessarily, which is consistent with results in previous emission-line studies \citep{Steidel:2016, Trainor.R:2016a, Shapley.A:2019a}.
We note that in our models the \NO\ ratio follows a single relation of Equation \ref{eq:6}, although observed \NO\ ratios have a large dispersion \citep{Pilyugin.L:2012a}.
Including the \NO\ dispersion in the models may help to explain high \([\OIII]/\Hb\) ratios \citep{Strom.A:2018a, Curti.M:2022a} by moving the model grids to high \([\NII]/\Ha\) direction (see Figure \ref{fig:bptfir_grid}).

In the FIR diagram, the \(\LOiii/\SFR\) ratios are closely related with input ionizing spectra.
Four of the \(z > 6\) galaxies---SXDF-NB1006-2, J0217-0208, J0235-0532, and J1211-0118---exhibit high \(\LOiii/\SFR\) ratios of \(\log{\LOiii/\SFR\ [\LoMoyr]} > 7.7\).
These \(\LOiii/\SFR\) ratios are too high to be explained by photoionization models adopting a spectrum of BPASS \(300\) Myr cSF model, \(Z_{*}/\Zgas = 1\), and \(1.5 < \log{n_\text{H}/\cmmm} < 3.0\).
Changing either assumption of the models is necessarily to explain the high \(\LOiii/\SFR\) ratios.

A possible solution is low \(Z_{*}/\Zgas\), which is adopted in our fiducial photoionization models.
Low \(Z_{*}/\Zgas\) increases ionizing photon flux illuminating metal-enriched gas per \SFR\ as well as hardness of ionizing spectra.
These result in high \(L_{\Hb}/\SFR\) and high \(\OIIIfir/\Hb\), respectively, and hence high \(\LOiii/\SFR\) ratios (see Appendix \ref{sec:sfr-conversion}).
Indeed, the fiducial models assume  \(Z_{*}/\Zgas = 0.2\) and successfully explain the \(z > 6\) galaxies with the high \(\LOiii/\SFR\) ratios within line measurement errors.
Although low \(Z_{*}/\Zgas\) at high redshift was inferred from UV-to-optical emission-line studies \citep{Cullen.F:2019a, Harikane.Y:2020a, Kashino.D:2022a}, our photoionization models highlight the necessity of low \(Z_{*}/\Zgas\) ratios at \(z > 6\) from FIR \(\LOiii/\SFR\) ratios.

Bursty or increasing star-formation history is an alternative solution to explain the high \(\LOiii/\SFR\) ratios.
For a part of the \(z > 6\) galaxies, short star-formation ages of \( \lesssim 5\) Myr are inferred from the SED fittings \citep{Inoue.A:2016a, Hashimoto.T:2018a, Tamura.Y:2019a, Hashimoto.T:2019a}.
Such a bursty or increasing star-formation history leads to underestimating \SFRUVIR\ and thus overestimating \(\LOiii/\SFRUVIR\), as illustrated in Figure \ref{fig:fir-name} where all the \SFRSED\ are higher than \SFRUVIR\ estimated by \citet{Carniani.S:2020a}.
Moreover, this solution is in line with theories that bursty star formation can explain high \(\OIIIfir/\CIIfir\) ratios of the \(z > 6\) galaxies \citep{Arata.S:2020a, Vallini.L:2021a}.
The bursty star-formation history is compatible with low \(Z_{*}/\Zgas\); therefore, the high \(\LOiii/\SFR\) ratios support both of these two properties regarding the ionizing spectrum for the \(z > 6\) galaxies.

Finally, we mention possible explanations of high \(\LOiii/\SFR\) ratios other than low \(Z_{*}/\Zgas\) and bursty star-formation history.
Photoionization models with \(\log{n_\text{H}/\cmmm} < 1.0\) can also explain the high \(\LOiii/\SFR\) ratios, as shown in \citet{Yang.S:2020a}.
However, this models require high \(U\) (\(\log U > -2.0\)) and \(\Zgas/\Zo \sim 1\) as well as low \(n_\text{H}\), which are totally different from the nebular parameters of the \(z\sim0\) and \(2\) galaxies and would contrast with predictions that high-redshift galaxies exhibit higher electron densities than galaxies at \( z\sim0 \) \citep[e.g.,][]{Sanders.R:2016a, Kaasinen.M:2017a}.
We note that highly-ionized diffuse gas (having high \(U\) and low \(n_\text{H}\)) may contribute \OIIIfir\ emission as observed in the local universe \citep{Kawada.M:2011a, Lebouteiller.V:2012a, Polles.F:2019a}.
High dust temperature is another possibility to be considered.
The dust temperature is highly uncertain for the \(z > 6\) galaxies because only one or two ALMA measurements are available in the most cases.
Higher dust temperature yields higher \(L_\text{FIR}\) and higher \SFRUVIR\ at given dust continuum fluxes.
If all the \(z > 6\) galaxies with the high \(\LOiii/\SFR\) have dust temperature of \(\gtrsim 60\) K \citep[][Ono et al.\! in prep.]{Harikane.Y:2020b}, \(\LOiii/\SFR\) becomes small enough to be explained by photoionization models.

One may expect that a top-heavy IMF can explain the high \(\LOiii/\SFR\) ratios of the \(z > 6\) galaxies.
To test this possibility, we adopted BPASS models using double power-law IMF with a massive slope of \(-2.0\) at \(0.5\)--\(100\) \Mo\ to construct a photoionization model.
This top-heavy model failed to reproduce the high \(\LOiii/\SFR\) ratios because predicted \(\OIIIfir/\Hb\) is similar to the Chabrier IMF models.

\subsection{Parameter dependence of photoionization models}
\label{sec:param-depend-photo}

\begin{figure}[t]
    \epsscale{1.2}
    \plotone{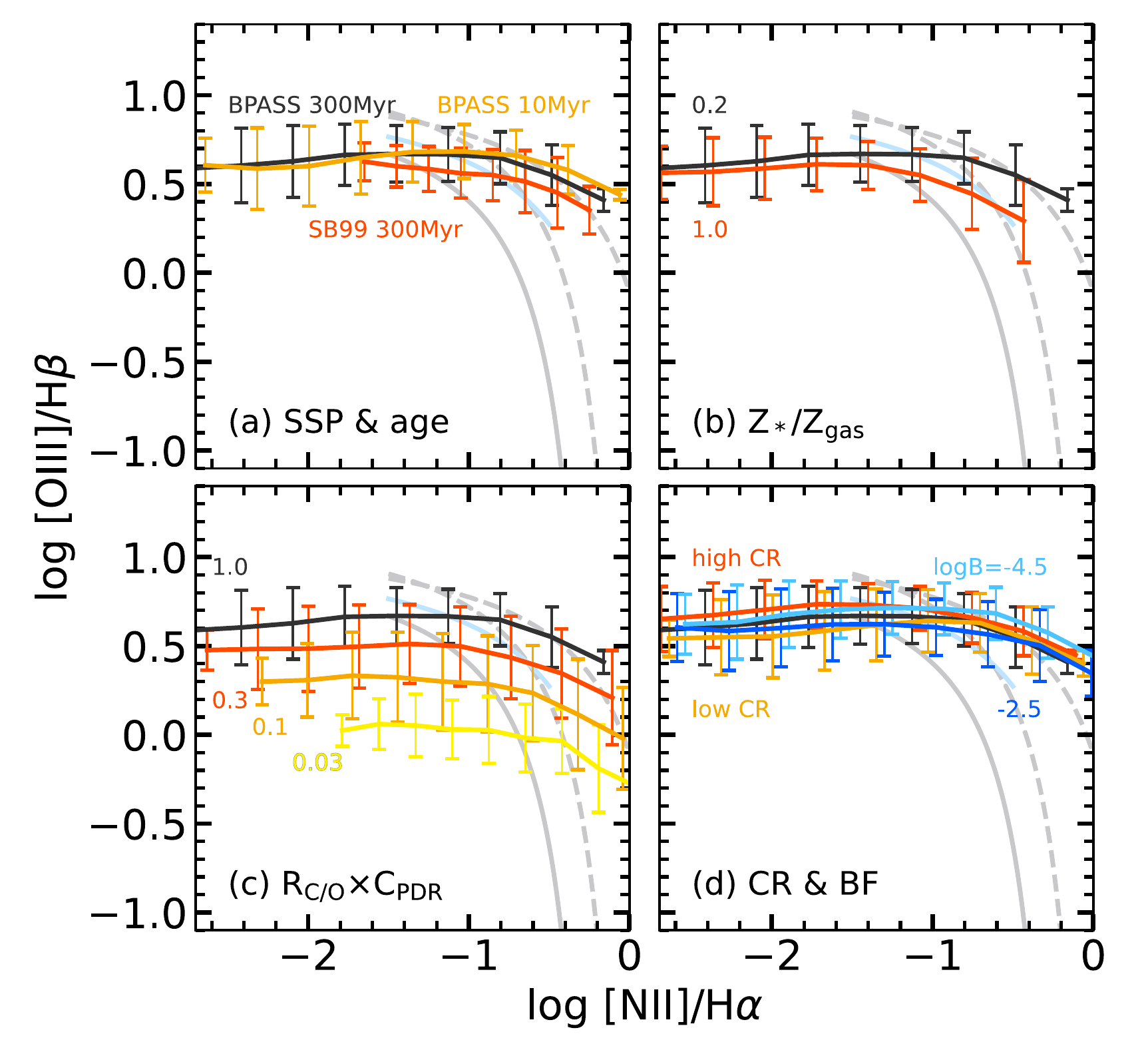}
    \caption{Photoionization model solutions on the BPT diagrams obtained for the \( z > 6 \) galaxies under the different assumptions (i.e., non-fiducial models).
      The obtained solutions are displayed with a line and bars, representing means and widths of the distribution at several \([\NII]/\Ha\) values, respectively.
      The fiducial model, which is the same as the left panel of Figure \ref{fig:result-fir}, is shown in black in all the panels.
      (a) Different SSP models and ages of the input ionization spectra: BPASS \( 300 \) Myr, SB99 \( 300 \) Myr, and BPASS \( 10 \) Myr constant star-formation models.
      The SB99 model only shows \(\log [\NII]/\Ha > -1.8\) (see text).
      (b) Different stellar-to-gaseous metallicity ratios: \(Z_{*}/Z_\text{gas} = 1.0\) and \(0.2\).
      (c) Different \CO\ abundance ratio and PDR covering fraction: \(R_{\CO}\CPDR = 1.0\), \(0.3\), \(0.1\), and \(0.03\).
      (d) Different cosmic-ray background intensity and strength of magnetic field: \(\times10\), \(\times1\), and \(\times0.1\), and \(\log{B}/\text{G} = -3.5\), \(-4.5\), and \(-5.5\).
      The dashed gray lines depict the criteria between star-forming galaxies and AGNs \citep{Kauffmann.G:2003a, Kewley.L:2001b}.
      Only low \CO\ and \CPDR\ increase a dispersion of \([\OIII]/\Hb\) while other assumptions do not affect the model predictions.
    }
  \label{fig:dis-parameter-dependence}
\end{figure}

Although we assumed several parameters in the fiducial photoionization models (Section \ref{sec:photo-ioniz-model}), the assumed parameters for the \( z > 6 \) galaxies are the same as for the \(z\sim2\) galaxies because most of them have not been constrained observationally.
In this section, we change parameters assumed for the \(z > 6\) galaxies in the fiducial models to assess uncertainties of our model predictions.
This section discusses the BPT diagram, but all the BPT, FIR, and \(U\nn\Zgas\) diagrams for the changed models are shown in Appendix \ref{sec:non-fiducial}.

\smallskip
\textit{SSP model and stellar age}---The fiducial photoionization models adopt BPASS \( 300 \) Myr cSF model.
We changed the input ionizing spectrum to SB99 \(300\) Myr cSF and BPASS \( 10 \) Myr cSF models.
The panel (a) of Figure \ref{fig:dis-parameter-dependence} compares predicted BPT diagrams for the \(z > 6\) galaxies.
We note that the panel (a) only shows a range of \(\log [\NII]/\Ha > -1.8\) for the SB99 model because the minimum \(Z_{*}\) in SB99 of \(0.05 \Zo\) prevents modeling at \(\Zgas/\Zo < 0.25\) using the models with \(Z_{*}/\Zgas = 0.2\).
Although there are slight differences at high \([\NII]/\Ha\) (i.e, high metallicity), all the models predict the same \([\OIII]/\Hb\) values at low \([\NII]/\Ha\) (low metallicity).
The differences of SSP models and stellar ages do not change the predicted BPT diagram for the \(z > 6\) galaxies.

\smallskip
\textit{Stellar-to-gaseous metallicity ratio}---The fiducial models for the \(z > 6\) galaxies assume \(Z_{*}/\Zgas = 0.2\) by following previous studies on galaxies at \(z\gtrsim2\).
We changed \(Z_{*}/\Zgas\) from \(0.2\) to \(1.0\).
In the panel (b) of Figure \ref{fig:dis-parameter-dependence}, the \(Z_{*}/\Zgas\) ratios do not affect predicted distributions at low \([\NII]/\Ha\) (i.e., low metallicity), while they slightly affect the distributions at \(\log{[\NII]/\Ha} > -1.5\) (high metallicity).
These results are similar to the case when SSP models and stellar ages change.

\smallskip
\textit{\CO\ abundance ratio and PDR covering fraction}---These two parameters change \CIIfir\ line intensity at a given metallicity.
The \CO\ ratio directly scales \(\OIIIfir/\CIIfir\) ratio.
We define \(R_{\CO} \equiv (\CO)/(\CO)_\text{ISM}\), where \((\CO)_\text{ISM}\) is the default \CO\ ratio of the ISM in the models, \(\log(\CO)_\text{ISM} = 0.02\).
Although we assumed \(R_{\CO} = 1\) irrespective of redshifts, observations of stars, local dwarf galaxies, and dumped \Lya\ systems suggest a metallicity dependence of \CO\ \citep{Penprase.B:2010a, Berg.D:2016a} and low-metallicity dwarf galaxies exhibit an average of \(\log(\CO) = -0.71\) (\(R_{\CO} = 0.19\)) at \(\log\Zgas/\Zsun \sim - 1\) \citep{Berg.D:2019a}.
\CPDR\ is a fraction at which an \HII\ region is covered by PDR.
This parameter is sometimes employed in photoionization models in the literature \citep[e.g.,][]{Cormier.D:2019a, Harikane.Y:2020b}.
Because [\CII] line intensity is almost linearly proportional to \(R_{\CO}\) and \CPDR\footnote{Actually, the [\CII] line emits from \HII\ regions even if \(\CPDR = 0\), so that our prescription simplifies \CPDR\ effects. In local dwarfs, a fraction of [\CII] emission from the \HII\ region is \( < 15\)\% to the total \citep{Cormier.D:2015a}.}, we changed \(R_{\CO} \CPDR\) to \(0.3\), \(0.1\), and \(0.03\) by reducing [\CII] intensity at those fractions.

In the panel (c) of Figure \ref{fig:dis-parameter-dependence}, the predicted \([\OIII]/\Hb\) value decreases as \(R_{\CO}\CPDR\) decreases.
Figure \ref{fig:appendix} in the appendix shows that the decrease in \([\OIII]/\Hb\) is attributed to a decrease in \(U\) at a given \Zgas.
Although the fiducial models explain high \(\OIIIfir/\CIIfir\) ratios of the \(z > 6\) galaxies by increasing \(U\) (Section \ref{sec:bridge-galpop}), low \CPDR\ and \(R_{\CO}\) can explain the high \(\OIIIfir/\CIIfir\) ratios with keeping \(U\) low \citep{Harikane.Y:2020b}.
However, Figure \ref{fig:appendix} also shows that it is more difficult for low \(R_{\CO}\CPDR\) models to explain galaxies with high \(\LOiii/\SFR\) than for the fiducial models.
For this reason, not all the \(z > 6\) galaxies would exhibit low \CO\ and \(\CPDR\).
If \(R_{\CO}\CPDR\) values vary among the \(z > 6\) galaxies, \([\OIII]/\Hb\) values have a dispersion of more than \( 0.5\) dex, larger than a prediction of the fiducial models.

\smallskip
\textit{Cosmic ray and magnetic field}---The background cosmic-ray intensity flux and the magnetic field amplitudes in the ISM is quite unclear at high redshift.
As an experiment, we changed the two parameters to \(1\)-dex higher and lower than the fiducial values.
The panel (d) of Figure \ref{fig:dis-parameter-dependence} illustrates that the two parameters change the predicted distributions by \( < 0.2\) dex in \([\OIII]/\Hb\) at most, which is much lower than the changes generated by \CO\ and \CPDR.
Therefore, the effects of the cosmic rays and magnetic fields are small despite their unclear amplitudes at high redshifts.

\smallskip
A prediction of our fiducial models, \(\log{[\OIII]/\Hb} \simeq 0.7\) for the \(z > 6\) galaxies, holds even when many assumed parameters change, while low \CO\ and \CPDR\ values make a dispersion of \([\OIII]/\Hb\) larger.
These predicted BPT diagrams will be tested by the upcoming \textit{JWST} observations.
\textit{JWST} GTO and GO cycle 1 programs plan to perform follow-up observations of the \(z > 6\) galaxies with the NIRSPEC integrated field units.
In the NIRSPEC observations, the dispersion of \([\OIII]/\Hb\) may be helpful to distinguish whether the ISM of the \(z > 6\) galaxies have low \CO\ and \CPDR.
The MIRI MIR observations targeting \Ha\ and [\NII] emission lines will also contribute to estimate the metallicity of the galaxies.

\subsection{Caveats}
\label{sec:caveats}

\smallskip
\textit{Aperture differences in optical and FIR observations}---The optical and FIR line fluxes used in this work were measured with different aperture sizes.
Optical spectroscopy usually uses slits or fibers, which can observe only a part of a nearby galaxy, whereas FIR instruments have large apertures to observe a whole galaxy.
These aperture differences may cause systematics in estimates of the nebular parameters from optical and FIR observations, even though our models assumed that the same nebular parameters can simultaneously explain galaxy distributions on the BPT and FIR diagrams.
We attempted to reduce the aperture differences by using the integrated spectroscopic observations as much as possible.
In the near future, the \(z > 6\) galaxies will be observed with \textit{JWST} integral field units (IFU) that covers the entire of them within its field of view.
Therefore, observing local galaxies with optical IFU will be important for a fair comparison with the \(z > 6\) galaxies.

\smallskip
\textit{Geometrical variation of dust attenuation}---This work ignores effects of the dust attenuation because FIR lines are not attenuated by dust and the BPT diagram takes ratios of lines with close wavelengths.
However, the geometrical variation of dust attenuation may bias optical and FIR radiation sources within galaxies.
Although we can observe FIR emission lines from a whole galaxy, the optical line fluxes may be dominated by emission from less dusty regions.
Thus, this spatial variation of attenuation may break our assumption that we can reproduce both optical and FIR emission-line ratios with the same nebular parameters.
Some studies report spatial offsets between UV and FIR line/continuum detections at \(z > 5\) \citep{Carniani.S:2018a, Bowler.R:2022a}.
These effects should be considered in the future analyses.

\smallskip
\textit{Contributions from DIG and AGN}---Our models ignore the DIG, which contributes low-ionization emission lines such as [\NII] and [\CII] lines in local galaxies \citep[e.g.,][]{Martin.C:1997a, Kaufman.M:2006a}.
The DIG increases the \([\NII]/\Ha\) ratio by a few times 0.1 dex and systematically biases metallicity estimates \citep{Zhang.K:2017a}, but its effect to \([\NII]/\Ha\) (and to the BPT diagram in this work) is relatively smaller than those to \([\SII]/\Ha\) and \([\OII]/\Hb\) \citep{Sanders.R:2017a}.
At \(z\sim2\), and possibly at higher redshift, DIG contributions to low ionization lines would be negligible because of high \SFR\ surface densities of galaxies at the redshift \citep{Sanders.R:2017a, Shapley.A:2019a}.
Our models also ignore the hidden AGN contributions.
Although we cannot reject a possibility that AGNs affects the line ratios of the \(z > 6\) galaxies, currently there are no clear signatures of AGNs in the \(z > 6\) galaxies.

\smallskip
\textit{Geometry of the ionized and photodissociated regions}---Most photoionization models, including ours, assume idealized nebular geometries such as spherical or plane-parallel geometry.
However, the actual \HII\ regions and galaxies have complex gaseous geometry, stellar distributions, and temperature/density structures.
This discrepancy between models and observations may complicate interpretations of observed emission-line ratios of galaxies.
More sophisticated photoionization modeling will be necessarily in the era of multi-wavelength, IFU observations \citep[e.g.,][]{Jin.Y:2022a}.

\section{Summary}
\label{sec:summary}
We have performed photoionization modeling to distributions of galaxy populations from \(z\sim0\) to \(z > 6\) on the BPT (\([\OIII]\lambda5007/\Hb\)--\([\NII]\lambda6585/\Ha\)) and FIR (\(\LOiii/\SFR\)--\(\LCii/\SFR\)) diagrams.
The galaxy samples are divided into three according to their available measurements: the optFIR sample (local dwarfs and U/LIRGs), the optical sample  (\(z\sim0\) and \(2\) galaxies), and the FIR sample (\(z > 6\) galaxies).
The constructed photoionization models have three free nebular parameters of the ionization parameter \(U\), hydrogen density \(n_\text{H}\), and gaseous metallicity \Zgas.
We have defined the regions representing distributions of the galaxy populations on the BPT and FIR diagrams and have searched nebular parameters that satisfy the regions.

Our fiducial photoionization models successfully reproduce the nebular parameters of the local dwarfs and U/LIRGs consistent with results of previous studies.
Then we have applied the photoionization models to the optical and FIR samples.
For the \(z\sim0\) galaxies, the predicted distribution on the FIR diagram is consistent with \textit{ISO} observations.
The \(z\sim2\) galaxies are predicted to exhibit higher \(\OIIIfir/\CIIfir\) ratios than the \(z\sim0\) galaxies on average.
For the \(z > 6\) galaxies, the predicted nebular parameters have a negative \(U\nn\Zgas\) correlation and \(U\) and \Zgas\ values are comparable to those of Lyman-alpha emitters at \(z\sim2\).
The distribution on the BPT diagram for the \(z > 6\) galaxies is relatively flat; \(\log{[\OIII]/\Hb} = 0.5\nn0.8\) while \([\NII]/\Ha\) strongly depends on \Zgas.

Comparing the galaxy populations illustrates continuous distributions from low- to high-mass galaxies at \(z\sim0\) and continuous shifts from \(z\sim0\) to \(z > 6\), on all the diagrams (Figure \ref{fig:result-summary}).
The \(z\sim0\) galaxies bridge the distributions of the local dwarfs and U/LIRGs on the FIR and \(U\nn\Zgas\) diagrams, as well as on the BPT diagram.
These diagrams show continuous transitions of the stellar and ISM properties from dwarfs through normal star-forming galaxies to U/LIRGs.
The \(z > 6\) galaxies have higher \(U\) than the \(z\sim0\) and \(2\) galaxies at given \Zgas.
Thanks to high \(U\) and assumed hard ionizing spectrum, the \(z > 6\) galaxies have higher \([\OIII]/\Hb\) on the BPT diagram and higher \(\OIIIfir/\CIIfir\) on the FIR diagram than the \(z\sim0\) and \(2\) galaxies.
These continuous shifts on the diagrams demonstrate the redshift evolution of the stellar and ISM properties of galaxies from the epoch of reionization to present.
In addition, the \(z > 6\) galaxies share a large part of the distributions with the local dwarfs, indicating a similarity of ISM properties between them.

We find that some of the \(z > 6\) galaxies exhibit high \(\LOiii/\SFR\) ratios, which cannot be reproduced by BPASS \(300\) Myr constant star-formation models with \(Z_{*}/\Zgas = 1\) and \(1.5 < n_\text{H}/\cmmm < 3.0\).
To reproduce such high \(\LOiii/\SFR\) ratios, our photoionization models require: 1) low stellar-to-gaseous metallicity ratios of \(Z_{*}/\Zgas = 0.2\), consistent with results of UV-to-optical emission line studies; or 2) bursty or increasing star-formation history, in agreement with the SED-fitting results of some \(z > 6\) galaxies.
We mentioned other possibilities of low hydrogen density and high dust temperature.
Similarly, high \([\OIII]/\Hb\) ratios of the local dwarfs and \(z\sim2\) galaxies require BPASS models with \(Z_{*}/\Zgas = 1\) and \(0.2\), respectively, rather than SB99 non-rotating models.

Upcoming \textit{JWST} plans to detect rest-frame optical emission lines of the \(z > 6\) galaxies.
These new NIR observations can test our model predictions, including the distributions on the BPT diagrams and the nebular parameters of the \(z > 6\) galaxies, and improve our understanding of the early universe.

\begin{acknowledgments}
We thank Hanae Inami for her helpful comments in conferences.
We wish to thank the anonymous referee for valuable comments to improve our manuscript.
This research is supported by NAOJ ALMA Scientific Research Grant number 2020-16B and by JSPS KAKENHI Grant Number 17H01114 and 21K13953.
TH was supported by Leading Initiative for Excellent Young Researchers, MEXT, Japan (HJH02007).
ALMA is a partnership of ESO (representing its member states), NSF (USA), and NINS (Japan), together with NRC (Canada), MOST and ASIAA (Taiwan), and KASI (Republic of Korea), in cooperation with the Republic of Chile.
The Joint ALMA Observatory is operated by ESO, AUI/NRAO, and NAOJ.
This research has made use of NASA’s Astrophysics Data System.
\end{acknowledgments}

\software{NumPy \citep{Harris.C:2020a}, SciPy \citep{Virtanen.P:2020a}, IPython \citep{Perez.F:2007a}, Matplotlib \citep{Hunter.J:2007a}, Astropy \citep{Astropy-Collaboration:2013a, Astropy-Collaboration:2018a}, \textsc{Cloudy} \citep{Ferland.G:2017a}}

\bibliographystyle{aasjournal}

\appendix
\restartappendixnumbering
\renewcommand{\theHfigure}{\thesection\arabic{figure}}

\section{The \(\lowercase{z} > 6\) galaxies used in this work}

\begin{figure}[t]
    \epsscale{1.2}
    \plotone{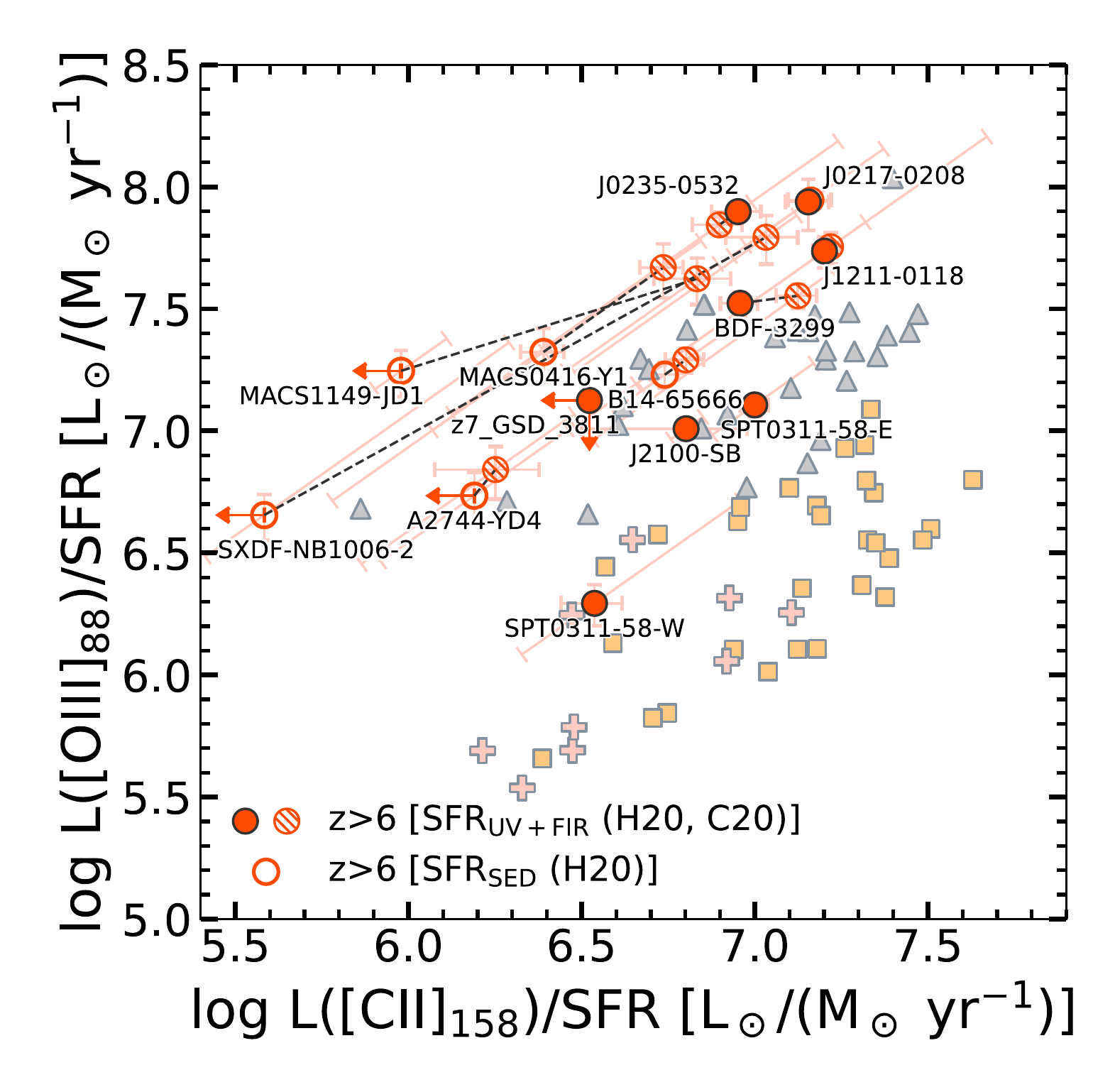}
    \caption{Same as the FIR diagram in Figure \ref{fig:bptfir}, but names of the \(z > 6\) galaxies are denoted near the data points.
      The black dashed lines connects two measurements for the same objects listed in \citet{Harikane.Y:2020b} and \citet{Carniani.S:2020a}.
    }
  \label{fig:fir-name}
\end{figure}

Figure \ref{fig:fir-name} shows names of the \(z > 6\) galaxies for convenience of the reader.
As nine UV-selected galaxies have two measurements given by \citet{Harikane.Y:2020b} and \citet{Carniani.S:2020a}, their two data points are connected with the dashed lines.

\section{\SFR\ conversion factors dependent on SSP models, metallicity and stellar age}
\label{sec:sfr-conversion}

\begin{figure}[t]
    \figurenum{B1}
    \epsscale{1.2}
    \plotone{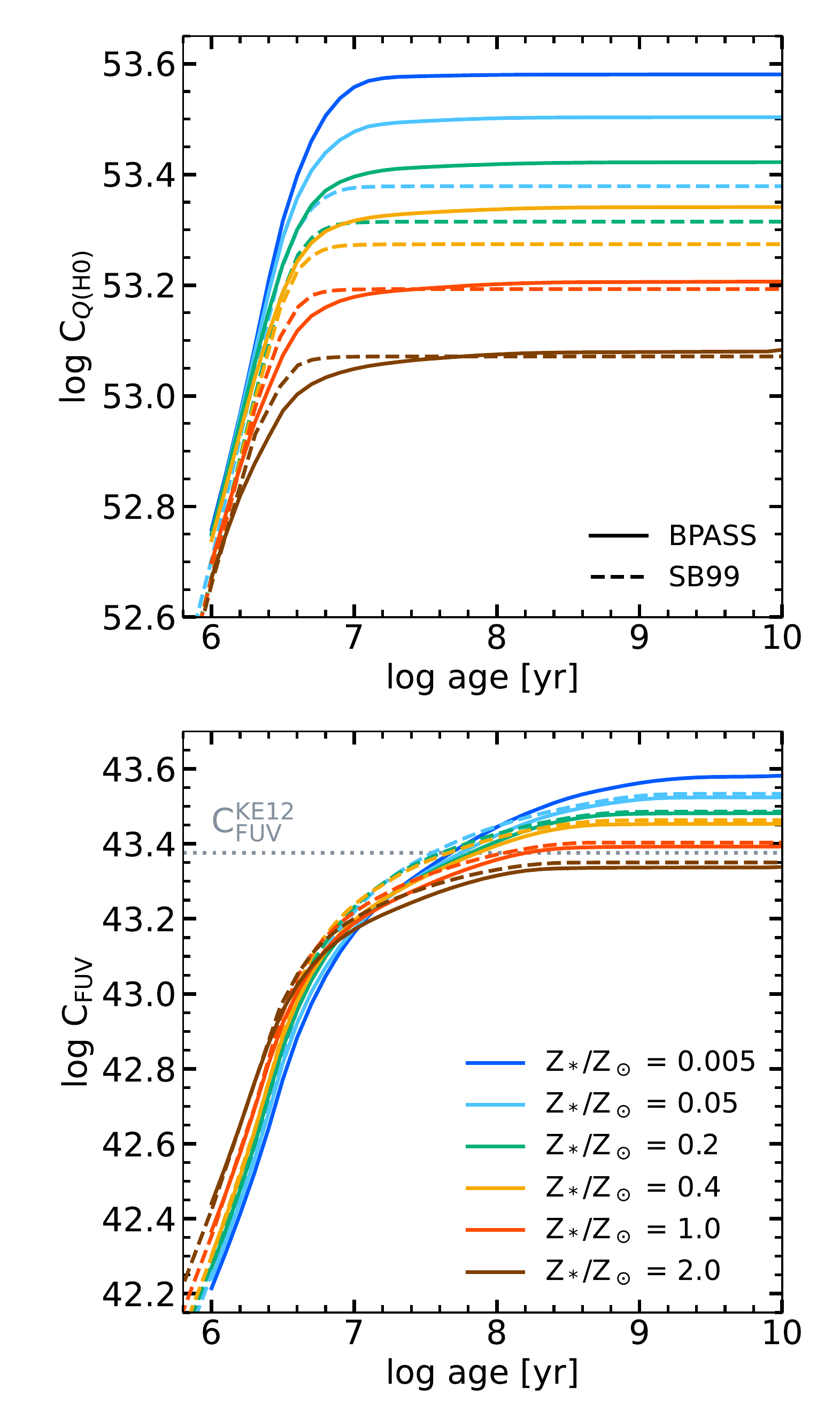}
    \caption{\SFR\ conversion factors as a function of the stellar metallicity and stellar age.
      The solid and dashed lines show the BPASS and SB99 SSP models.
      The line colors depend on metallicity.
      \textit{Top}: conversion factor for \(Q(H^0)\).
      \textit{Bottom}: conversion factor for \(L_\text{FUV}\).
      The gray dotted line indicates the conversion factor in \citet{Kennicutt.R:2012a}.
    }
  \label{fig:sfr-conv}
\end{figure}

Relations between \SFR\ and its tracers depend on the SSP models, star-formation history, metallicity, and IMF.
Even under constant star-formation history as assumed in this work, the relation depends on stellar population age; namely,
\begin{equation}
    \label{eq:2}
    L_x/\SFR = C_x (\text{SSP}, Z_{*}, t, \text{IMF}),
\end{equation}
where \(x\) denotes the tracer of \SFR, \(L_x\) the tracer luminosity, \(C_x\) the conversion factor, \(t\) the age.
Although most of the references cited in this work \citep{De-Looze.I:2014a, Madden.S:2013a, Howell.J:2010a, Herrera-Camus.R:2018a, Carniani.S:2020a, Harikane.Y:2020b} derived \SFR\ using formulae of \citet{Kennicutt.R:1998a}, \citet{Hao.C:2011a}, and \citet{Murphy.E:2011a} \citep[see][for a review]{Kennicutt.R:2012a}, these formulae basically assume SB99 SSP, \(100\) Myr cSF, and solar metallicity.
The effects of different assumptions have been discussed in, for example, \citet{Inoue.A:2011b} and \citet{Wilkins.S:2019a}.
This section revisits the \SFR\ conversion factors to make them useful for this work.

The production rate of ionizing photons is proportional to the starformation rate,
\begin{equation}
\label{eq:4}
\left( \frac{Q(H^0)/\mathrm{s^{-1}}}{\SFR/\Moyr} \right) = C_{Q(H^0)}(\text{SSP}, Z_{*}, t, \text{IMF}),
\end{equation}
where \(C_{Q(H^0)}\) can be determined from the SSP models.
The top panel of Figure \ref{fig:sfr-conv} shows the dependence of \(C_{Q(H^0)}\) on the metallicity and age for the SB99 and BPASS models.
The IMF is fixed to the \citet{Chabrier:2003} IMF because the different IMF changes the conversion factors only by a factor of a constant, among the \citet{Salpeter:1955}, \citet{Chabrier:2003}, and \citet{Kroupa:2001} IMF.

For the Case B recombination in an isothermal \HII\ region, \(Q(H^0)\) can be converted to the \Hb\ luminosity as
\begin{equation}
\label{eq:5}
L_{\Hb} = \frac{\gamma_{\Hb}(T_\text{e}, n_\text{e})}{\alpha_\text{B}(H^0, T_\text{e}, n_\text{e})}Q(H^0),
\end{equation}
where \(\gamma_{\Hb}\) is the \Hb\ emission coefficient and \(\alpha_\text{B}\) is the Case B recombination rate.
This assumption holds in the \textsc{Cloudy} calculations well.
The coefficients \(\gamma_{\Hb}\) and \(\alpha_\text{B}\) depend on the electron temperature \(T_\text{e}\) and electron density \(n_\text{e}\), but the fraction \(\gamma_{\Hb}/\alpha_\text{B}\) is almost independent of them.
We adopt \(\gamma_{\Hb}/\alpha_\text{B} = 4.65\times10^{-13}\) erg, uncertainties of which are within \(0.02\) dex at \(T_\text{e} = 5\nn20\times10^4\) K and \(n_\text{e} < 10^4\) \cmmm\ \citep{Storey.P:1995a}.
Hence, the relation between \SFR\ and \(L_{\Hb}\) is
\begin{align}
\label{eq:7}
  \left( \frac{L_{\Hb}/\ergs}{\SFR/\Moyr} \right) &= C_{\Hb} \nonumber \\
  &= 4.65\times10^{-13} ~ C_{Q(H^0)}.
\end{align}

\begin{figure}[t]
    \epsscale{1.2}
    \plotone{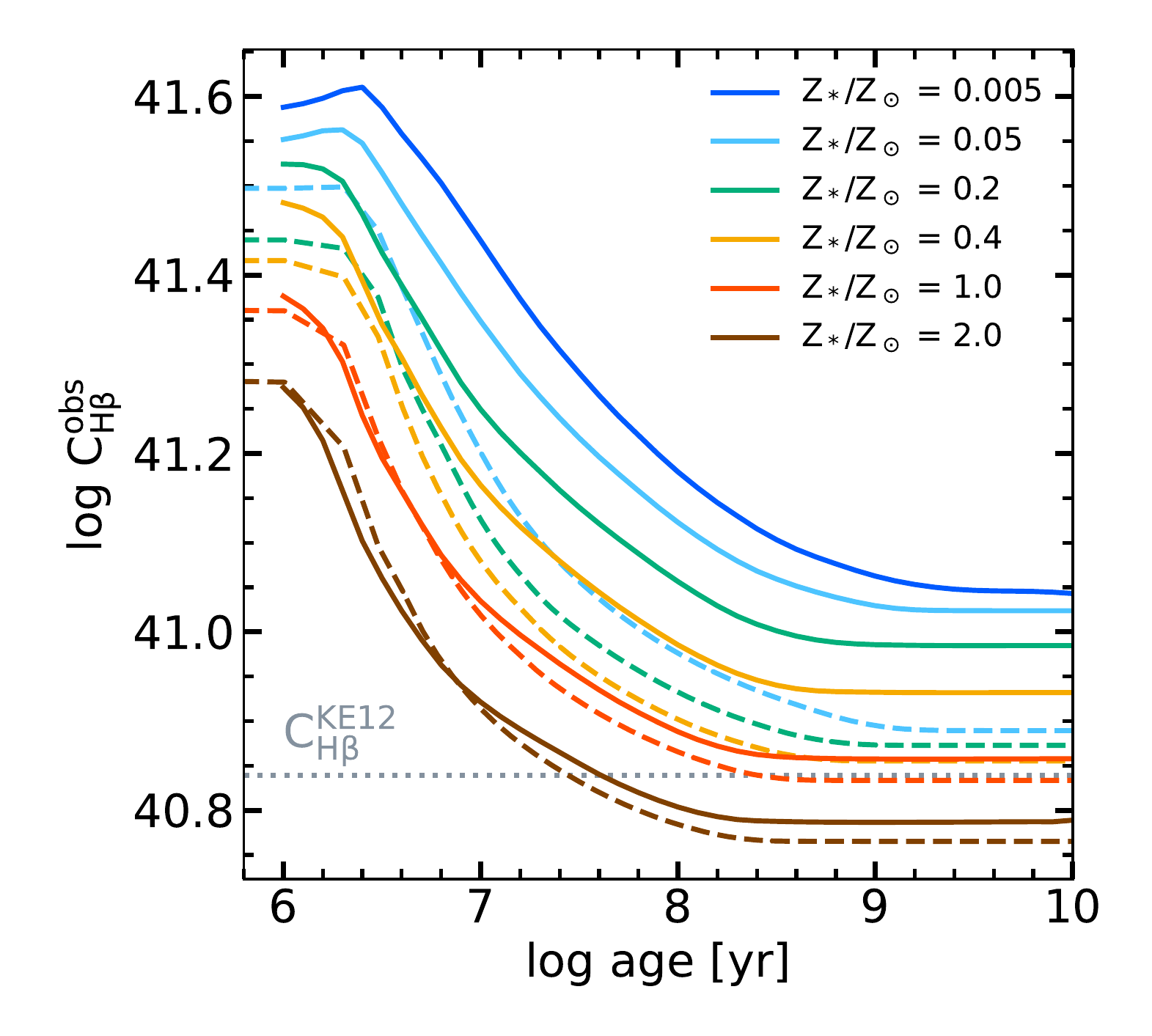}
    \caption{\SFR\ conversion factor for \(L_{\Hb}\) (Equation \ref{eq:10}) as a function of the stellar metallicity and stellar age.
     The solid and dashed lines show the BPASS and SB99 SSP models.
     The gray dotted line indicates the conversion factor in \citet{Kennicutt.R:2012a}, which is obtained by dividing \(C_{\Ha}^\text{KE12}\) by the \(\Hb/\Ha\) ratio of \(2.87\) under Case B recombination.
     This factor \(C_{\Hb}^\text{obs}\) converts \Hb\ intensity in models to \SFR\ that can be compared with observations.
    }
  \label{fig:sfr-conv-Hb}
\end{figure}

FUV luminosity is another tracer of \SFR.
We use the FUV luminosity of \(L_\text{FUV} = \lambda L_{\lambda}\) at \(\lambda = 1450\nn1520\) \AA\ to derive the conversion relation from the SSP models as
\begin{equation}
\label{eq:8}
\left( \frac{L_\text{FUV}/\ergs}{\SFR/\Moyr} \right) = C_{\text{FUV}}.
\end{equation}
The bottom panel of Figure \ref{fig:sfr-conv} shows the dependence of \(C_{\text{FUV}}\) on the parameters.

Equation \ref{eq:7} is available to convert line-to-\Hb\ ratios output by \textsc{Cloudy} to line-to-\SFR\ ratios; However, this \SFR\ is inconsistent with those estimated from observations used in this work because \(C_{Q(H^0)}\) depends on SSP, \(Z_{*}\), and \(t\).
In the observations, \SFR\ is often derived from the UV (and FIR) luminosity, following
\begin{equation}
\label{eq:9}
\left( \frac{L_\text{FUV}/\ergs}{\SFR^\text{obs}/\Moyr}  \right) = C^{\text{KE12}}_{\text{FUV}},
\end{equation}
where \(C^{\text{KE12}}_{\text{FUV}}\) is the conversion factor in \citet{Kennicutt.R:2012a}.
In \textsc{Cloudy} calculations, combining Equations \ref{eq:7} and \ref{eq:8} yields the conversion from \(L_{\text{FUV}}\) to \(L_{\Hb}\).
By inserting this conversion to Equation \ref{eq:9}, we obtain \(C_{\Hb}^\text{obs}\) as
\begin{equation}
\label{eq:10}
\left( \frac{L_{\Hb}/\ergs}{\SFR^\text{obs}/\Moyr}  \right) = \frac{C_{\Hb}(SSP, Z_{*}, t) ~ C^{\text{KE12}}_{\text{FUV}}}{C_{\text{FUV}}(SSP, Z_{*}, t)},
\end{equation}
which connect the \Hb\ luminosity in \textsc{Cloudy} to \SFR\ that can be compared with the observations.
Figure \ref{fig:sfr-conv-Hb} illustrates dependence of \(C_{\Hb}^\text{obs}\) on stellar age and metallicity.
\Hb\ luminosity at given \SFR\ increases as the stellar age and metallicity decrease.
Regarding the \(z > 6\) galaxies, while the observed \SFRUVIR\ can be compared with \(\SFR^\text{obs}\) of Equation \ref{eq:10}, the observed \SFRSED\ can be compared with \SFR\ of Equation \ref{eq:7} because the SED fitting already took stellar age and metallicity into account.

\section{Results of Non-Fiducial Photoionization Modeling}
\label{sec:non-fiducial}
In Section \ref{sec:param-depend-photo}, we changed assumptions of the photoionization models from the fiducial ones.
Figure \ref{fig:appendix} shows panels that are the same as Figure \ref{fig:result-fir} but for photoionization models with different assumptions.

\begin{figure*}[tbhp]
    \figurenum{C1}
    \epsscale{1.2}
    \plotone{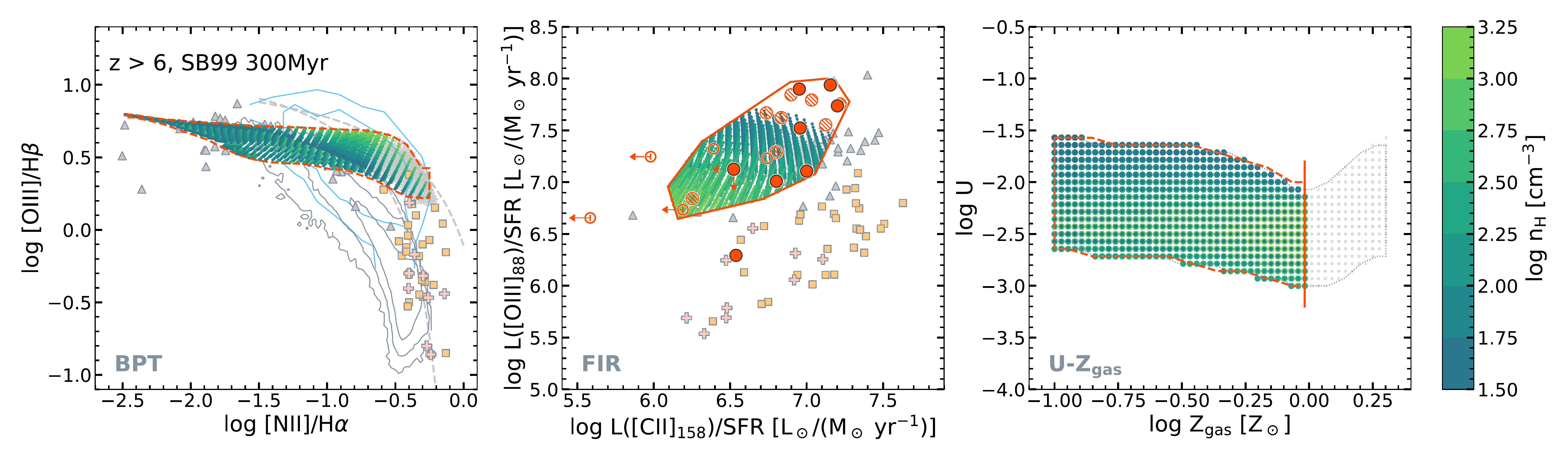}
    \plotone{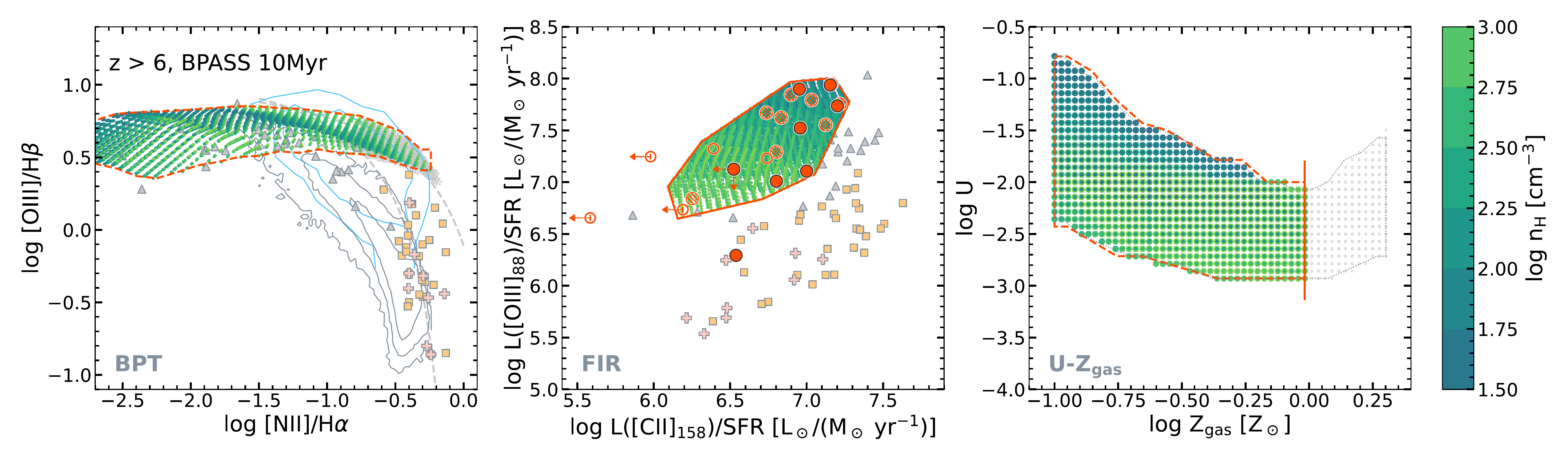}
    \plotone{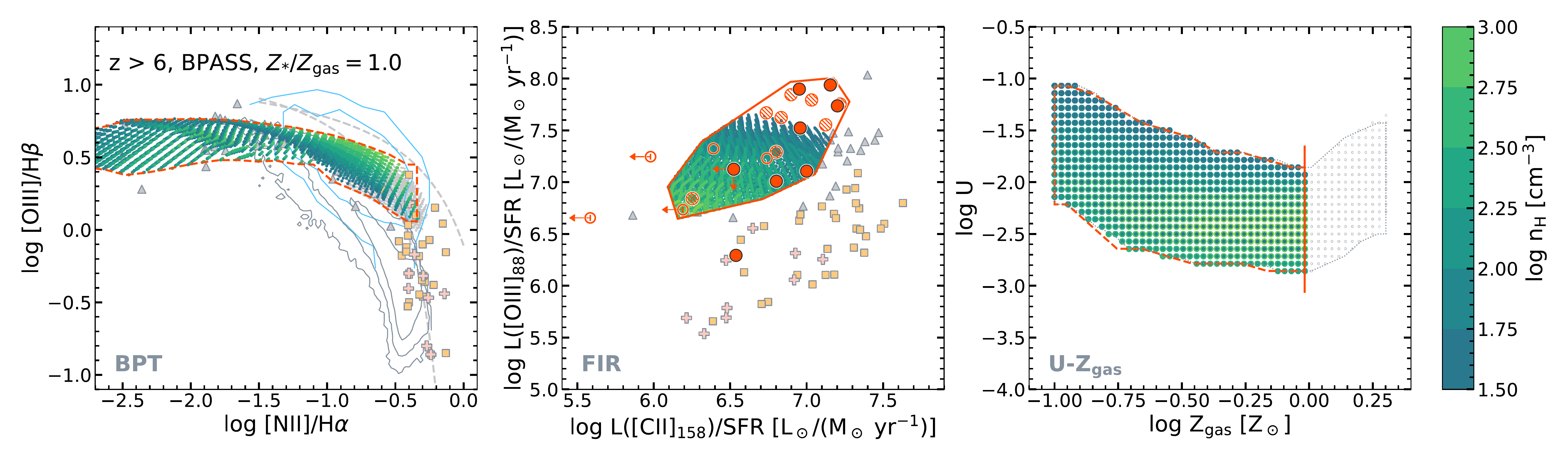}
    \caption{Same as Figure \ref{fig:result-fir}, but for photoionization models with assumptions that differs from the fiducial ones.
      The changed assumptions are written in the top left of the left panels.
    }
  \label{fig:appendix}
\end{figure*}

\begin{figure*}[t]
    \figurenum{C1}
    \epsscale{1.2}
    \plotone{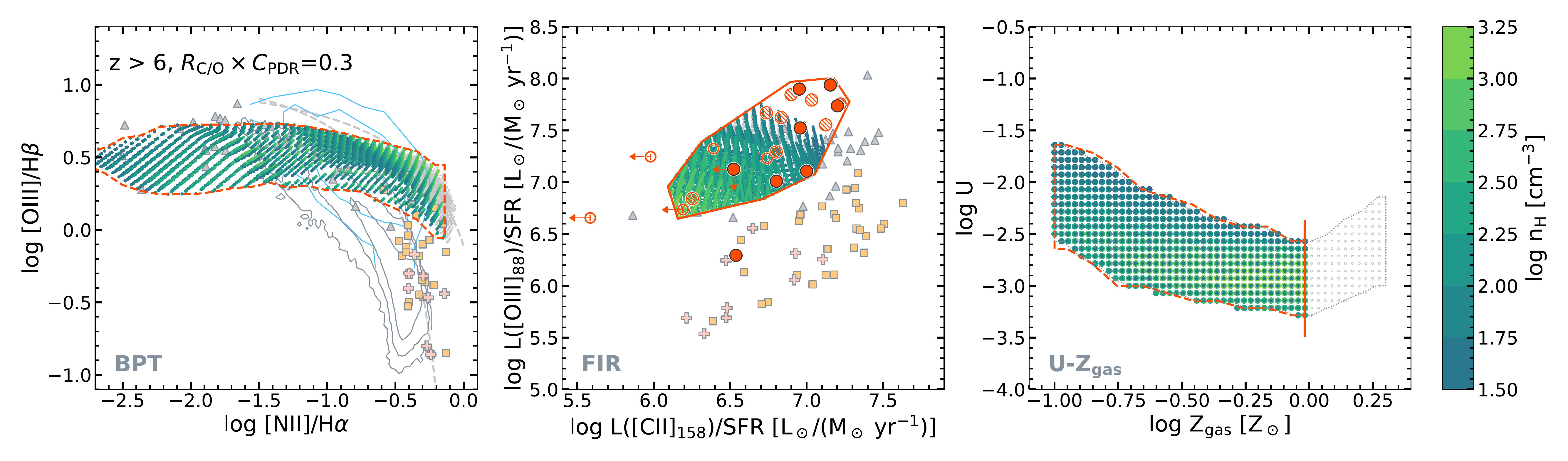}
    \plotone{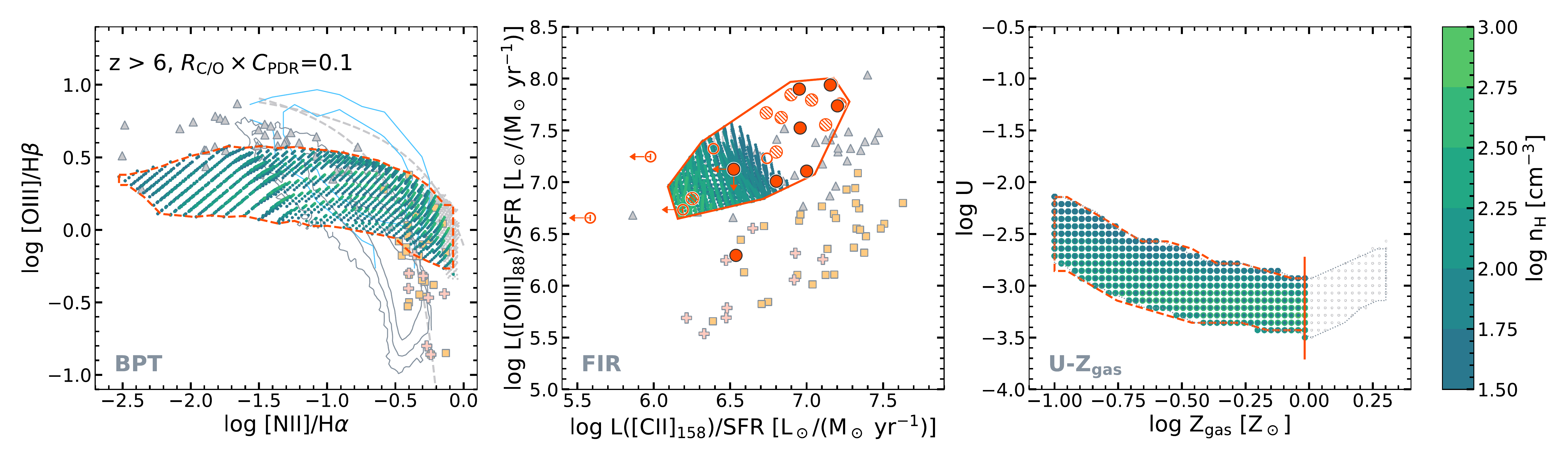}
    \plotone{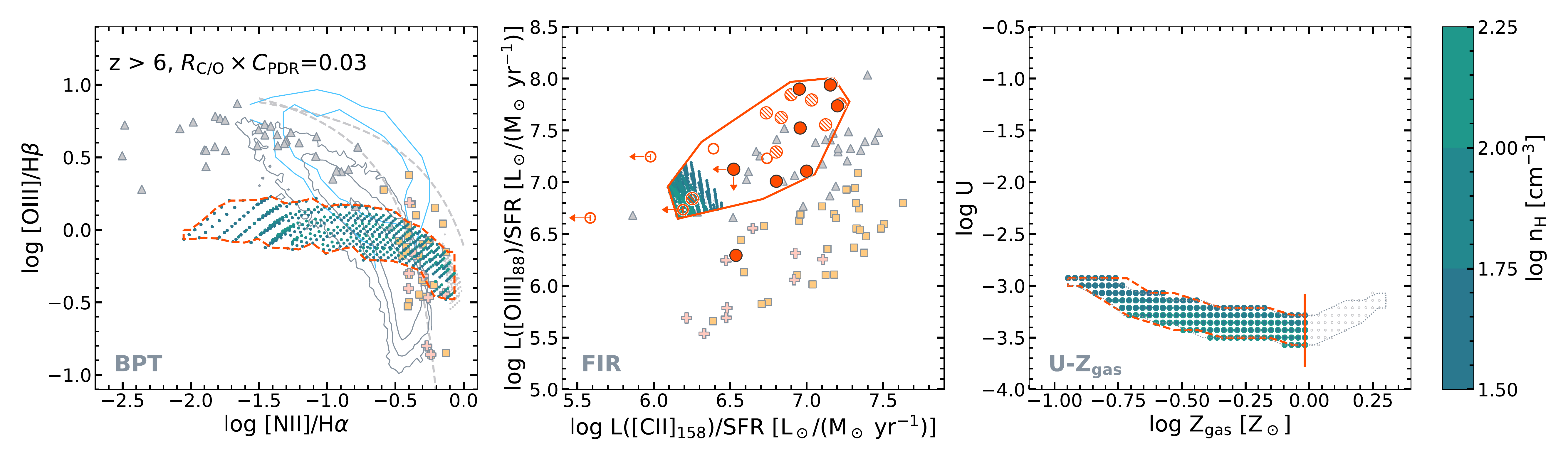}
    \caption{\textit{Continued.}
    }
\end{figure*}

\begin{figure*}[t]
    \figurenum{C1}
    \epsscale{1.2}
    \plotone{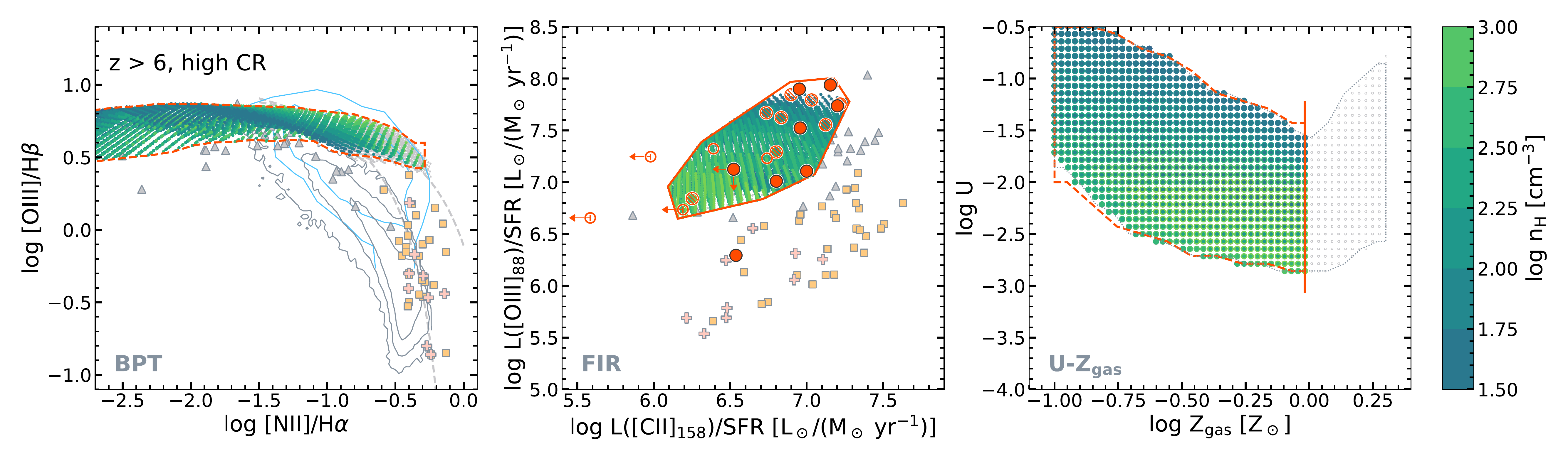}
    \plotone{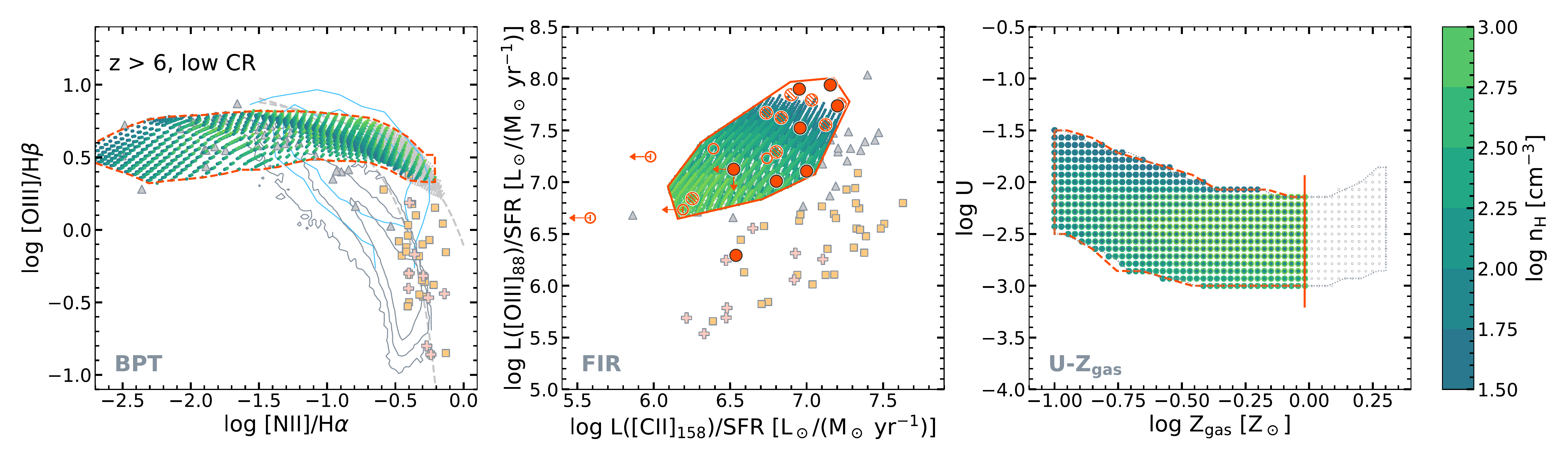}
    \plotone{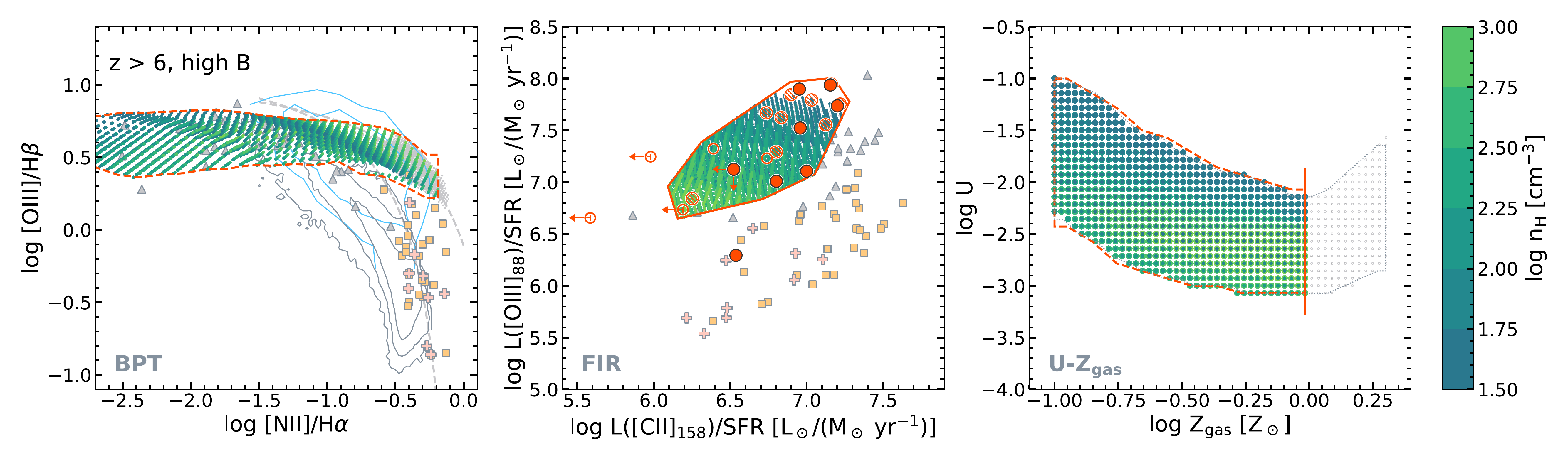}
    \plotone{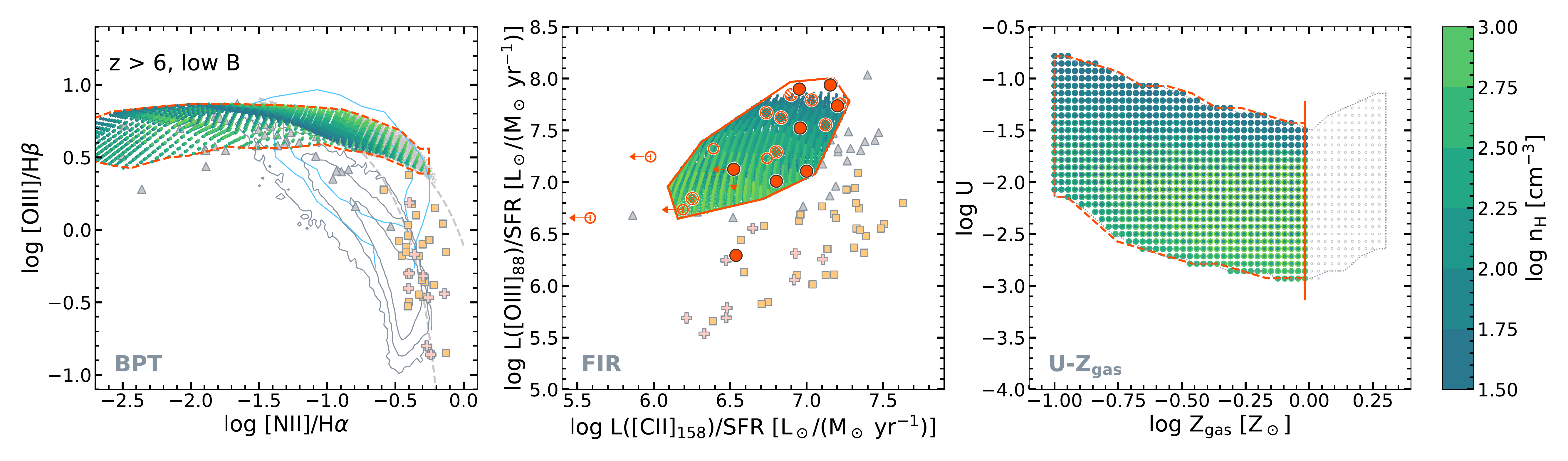}
    \caption{\textit{Continued.}
    }
\end{figure*}

\end{document}